\documentclass[a4paper,11pt,reqno]{article}

\usepackage{a4}
\usepackage{fullpage}

\usepackage{color}
\def\colzwo{} 

\usepackage[latin1]{inputenc}
\usepackage[english]{babel}
\usepackage{graphicx}
\usepackage{amsmath}
\usepackage{amsthm}
\usepackage{amssymb}
\usepackage{bbm}
\usepackage{url}
\usepackage{enumerate}

\usepackage{psfrag}
\usepackage{tikz}
\usepackage[centerlast,loose]{subfigure}
\usepackage{listings}

\theoremstyle{plain}

\theoremstyle{definition}

\theoremstyle{remark}

\newcommand{\N}{\ensuremath{\mathbb{N}}}

\renewcommand{\P}{\ensuremath{\mathbb{P}}}
\newcommand{\E}{\ensuremath{\mathbb{E}}}

\begin{document}

\title{Analysis of DNA sequence variation within marine species using Beta-coalescents}

\author{Matthias Steinr\"ucken\footnote{Department of Statistics, University of California, 367 Evans Hall MC 3860, Berkeley, CA 94720-3860, USA, e-mail: {\tt steinrue@stat.berkeley.edu} (corresponding author)}\ , Matthias Birkner\footnote{Johannes-Gutenberg-Universit\"{a}t Mainz, Institut f\"{u}r Mathematik, Staudingerweg 9, 55099 Mainz, Germany, e-mail: {\tt birkner@mathematik.uni-mainz.de}} \hspace{0em} and Jochen Blath\footnote{Technische Universit\"{a}t Berlin, Institut f\"{u}r Mathematik, Strasse des 17.\ Juni 136, 10623 Berlin, Germany, e-mail: {\tt blath@math.tu-berlin.de}}}

\date{}

\maketitle

\begin{abstract}
We apply recently developed inference methods based on general
coalescent processes to DNA sequence data
obtained from various marine species. Several of these species are believed
to exhibit so-called shallow gene genealogies, potentially due to extreme reproductive behaviour,
e.g.~{\colzwo via} Hedgecock's ``reproduction sweepstakes''.
Besides the data analysis, {\colzwo in particular the inference of 
mutation rates and the estimation of the (real) time to the
most recent common ancestor},
we briefly address the question whether the genealogies might be 
adequately described by so-called Beta coalescents (as opposed to Kingman's coalescent), 
allowing multiple mergers of genealogies.

{\colzwo The choice of the underlying coalescent model for the genealogy has drastic implications for the estimation of the above quantities, in particular the real-time embedding of the genealogy.}
\end{abstract}

AMS (2000) subject classification.
{\em Primary:}
62F99	
{\em Secondary:}
62P10;	
92D10;	
92D20   

\vspace{2mm}


\vspace{2mm}

\section{Introduction} 
\label{introduc}

Within the last decade, considerable attention has been turned to the
explanation of the fact that intra-species DNA sequence variation
yields shallow gene genealogies resp.~low ratio between effective
population size $N_e$ and adult census size $N$ in several marine
species (see, e.g., \cite{A04}, \cite{H94},
\cite{H05}, \cite{TWG02}, \cite{WT03}), as well as to the theory of
mathematical models which may describe such genealogies in terms of
so-called $\Lambda$-coalescents (see, e.g., \cite{P99}, \cite{S99}, \cite{DK99},
\cite{MS01}, \cite{EW06}, \cite{BB08a}).

Indeed, Hedgecock \cite{H94} proposes a mechanism to describe
substantial variation of reproductive success in marine species in
terms of so-called ``reproduction sweepstakes'', to be won by highly
successful individuals within each generation, i.e, a few or even a single individual
may replace a large fraction of the entire population. This requires great
individual fecundity and high mortality early in life as well as 
``sweepstake-like chances of matching reproductive activity with
oceanographic conditions conducive to 
gamete maturation, fertilization, larval 
development, settlement and successful recruitment to the adult spawning 
population''. 
As a result, Hedgecock claims that the variance in offspring numbers may be
orders of magnitudes higher than what standard binomial or Poisson
models predict, leading to a small $N_e/N$ ratio.
See also \cite{HP11} for a recent overview. 

However, if one tries to turn such a mechanism into a mathematical
population model, it turns out that in order to make sense for
large populations, sweepstakes whose size are a positive fraction of
the present population cannot be too frequent, in fact, the probability of such an event in a given generation must approach zero for large population sizes. Otherwise, the model would predict vanishing genetic variability, in contrast to the empirical observations (as already 
suggested by \'Arnason in \cite{A04}).

{\colzwo To circumvent such trivialities, we first} discuss two
rigorous mathematical population models in the classical ``Cannings' framework'' (\cite{C74}, \cite{C75})
that incorporate extreme reproductive events (due to Eldon and Wakeley
\cite{EW06} and Schweinsberg \cite{S03}), which can be considered as simple models of
Hedgecock's sweepstakes, but still lead to non-vanishing variability.  
A classification result due to M\"ohle and Sagitov (\cite{MS01}) then yields the 
required timescales for large population size, in a way that sweepstakes 
neither dominate (leading to vanishing genetic variability) nor become negligible.

It is important to note that the required time-scaling is mostly
``non-classical'' 
(i.e., unlike the Wright-Fisher model and its relatives, not a 
linear function of the model's census population size)
hence will also affect the scaling of the mutation rates and make the
concept of the so-called (Kingman-) coalescent effective population
size discussed in \cite{SKK+05} void, since the existence 
of the latter depends on a linear change in time-scale.

The resulting limiting ancestral processes embedded in 
our population models (with extreme reproduction due to 
sweepstake-like behaviour) coincide with special cases of the so-called
$\Lambda$-coalescents, i.e.~exchangeable coalescents, which allow
multiple collisions of lineages (extending the merely binary
collisions in the Kingman-coalescent setup), but not simultaneous multiple
collisions.  Such processes were introduced and studied by Pitman
\cite{P99} and Sagitov \cite{S99} and include the classical
Kingman-coalescent as a special case. However, 
the class of $\Lambda$-coalescents is vast, 
in particular allowing for any type of
probability distributions on the set of (random) sweepstakes sizes.
An important pair of questions in each concrete scenario therefore is:
{\em What is the ``right'' distribution on sweepstakes sizes, what
  is the right timescale?}
In \cite{EW06}, Eldon and Wakeley discuss a simple model, in which
sweepstake sizes are always the same fixed positive fraction of the
population size. They then fit their model, using a maximum-likelihood
method based on the number of segregating sites and total number of
mutations, 
to mitochondrial data from Pacific Oysters {\colzwo ({\em Crassostrea gigas})}, taken
from \cite{BBB94}, with the result that the maximum-likelihood 
estimator for the fixed sweepstakes size is $8\%$ of the living population.

To our knowledge, this {\colzwo was} the first time that a $\Lambda$-coalescent
based model has been calibrated to real data.  However, there are a
few issues that should be discussed.  From the modeling perspective,
there is no reason why there should be a fixed sweepstakes size.
Still, it is certainly infeasible to infer from the full
(non-parametric) class of $\Lambda$-coalescents.  Parametric
subclasses, which describe ``realistic'' mechanisms, would therefore
be of interest (we use certain so-called ``Beta-coalescents'', cf.~e.g.~\cite{BB08a}, 
\cite{BBC+05}, see Section~\ref{subset:exceptgenealog} and Section~\ref{ssn:model}
for a discussion of this class of coalescents).  Another important point is the adequacy of the equilibrium population assumption underlying
the coalescent model used in \cite{EW06}.
As pointed out in \cite{BBB94}, 
the pacific oyster data are taken from a population which
had only recently been introduced from {\colzwo Japan to Canada}. The shallow genealogy might 
therefore also be explained by the presence of a relatively recent
population bottleneck.

In the present article, we analyse several datasets {\colzwo obtained from} Atlantic Cod ({\em Gadus morhua}), 
taken from geographically separated locations, under the Beta-coalescent model, taking
the full information provided by the infinitely-many sites model into account (as opposed to \cite{EW06} where the authors use summary statistics based only on
the number of segregating sites and the total number of mutations).

As a result we report that in many cases, a neutral panmictic Kingman-based scenario can be rejected.
Further, our maximum-likelihood estimators for the parameters of the Beta-coalescents and mutation rates 
are presented, {\colzwo as well as an estimated real-time embedding of the genealogy and in particular the expected time to the most recent common ancestor given the data}. We finally discuss the question whether 
there is evidence for a sweepstake-based scenario in these datasets and 
propose and calibrate a potential candidate for the distribution of sweepstake sizes.



\section{Methods}
\label{Methods}

\subsection{Exceptional genealogies and exchangeable coalescents}
\label{subset:exceptgenealog}

Since the early 80ies, models based on the Kingman coalescent 
have been successfully used to describe the genealogy of
many biological populations. One of their distinguishing features is, together with exchangeability,
that only binary collisions are allowed.
That is, at most two ancestral lineages may coalesce at a time (exchangeability 
meaning that all pairs of lineages are treated equal).

However, it turns out that many species seem to exhibit ``exceptional genealogies'', 
for example ``shallow genealogies'', (cf.~e.g., \cite{A04}, \cite{H94}, \cite{H05}, \cite{TWG02}, \cite{WT03}),
which may be appropriately described by more general exchangeable coalescents, 
the so-called $\Lambda$-coalescents, which allow
{\em multiple collisions} of ancestral lineages. Indeed, under a Lambda-coalescent, given a sample of size $n$,
each $k$-tuple of ancestral lineages (where $2 \le k \le n$) is merging to form a single lineage 
at rate $\lambda_{n,k}$, where
\begin{equation}
\label{Lambdarates}
\lambda_{n, k} = \int_{[0,1]} x^{k}(1-x)^{n-k} \frac{1}{x^2} \Lambda(dx),
\end{equation}
for some finite measure $\Lambda$ on the unit interval $[0,1]$, see \cite{P99} {\colzwo or \cite{S99}}. 
Note that the family of Lambda-coalescents is rather large, and in
particular cannot be parametrised by finitely-many real variables. Important
examples include $\Lambda=\delta_0$ (Kingman's coalescent) and
$\Lambda=\delta_1$ (star-shaped genealogies). Here, we denote by 
$\delta_y$ the probability measure on $[0,1]$ with a unit point mass in $y \in [0,1]$. Note that this means that (for continuous functions $f$)
\begin{equation}
\label{eq:pointmass}
\int f(x)\, \delta_y(dx) = f(y).
\end{equation} 
From (\ref{Lambdarates}) and \eqref{eq:pointmass}, one 
readily obtains
\begin{align*}
\lambda_{n,k} = \int_{[0,1]} x^{k-2} (1-x)^{n-k} \delta_0(dx)
= 0^{k-2} =
\begin{cases}
1, & k=2,\\
0, & k >2,
\end{cases}
\end{align*}
in the Kingman case and 
\begin{align*}
\lambda_{n,k} = \int_{[0,1]} x^{k-2} (1-x)^{n-k} \delta_1(dx)
=
\begin{cases}
1, & k=n,\\
0, & 2 \le k < n,
\end{cases}
\end{align*}
in the star-shaped case. A little more generally, 
a single point mass at $\psi \in (0,1)$, say
$\Lambda=c \psi^2\delta_\psi$, $c>0$, yields 
\begin{align}\label{eq:EWatom}
\lambda_{n,k} &= 
c \int_{[0,1]} x^k (1-x)^{n-k} \frac{\psi^2}{x^2} \delta_\psi(dx) 
= \psi^k(1-\psi)^{n-k}.
\end{align}
This has a simple interpretation via independent Bernoulli trials 
with individual success probability $\psi$ that determines which 
lineages participate in a given merger event. Also, forwards in time,  
it has a simple interpretation for the
corresponding population model, namely, that on a macroscopic
time-scale, at rate $c>0$ a fraction of $\psi \cdot 100 \%$ of the living
population is replaced by the offspring of single ancestor 
(by the strong law of large numbers).
Such macroscopic reproduction events will be called ``extreme
reproductive behaviour'', and lead to multiple collisions in the
genealogy.  See, e.g., \cite{BB08b} for further details and
references.
\smallskip

Of course, from a modelling point of view a restriction to a fixed 
fraction $\psi$ appears unnatural, and the general case can be 
interpreted as a mixture over $\psi$'s. There are of 
course many possibility, and 
later, we will mainly be concerned with so-called {\em Beta-coalescents}, 
in particular the case where $\Lambda$ has a
Beta$(2-\alpha, \alpha)$-density, i.e.\ 
\[
\frac{\Gamma(2)}{\Gamma(\alpha)\Gamma(2-\alpha)} x^{1-\alpha}(1-x)^{\alpha-1}, 
\quad x \in (0,1)
\]
for some $\alpha \in (1,2)$ (where $\Gamma$ denotes Euler's Gamma function).  Even though this is not integrable for
$\alpha=2$, the transition rates then correspond to the classical
Kingman coalescent, which can in this way be included into the
Beta$(2-\alpha, \alpha)$-class, and intuitively smaller $\alpha$ 
corresponds to stronger skew in offspring distributions. 
This class of multiple merger coalescents 
is mathematically distinguished (revealing a close connection to 
$\alpha$-stable branching processes, see \cite{BBC+05}); 
furthermore, many large-sample properties are determined by the 
shape of $\Lambda$ near $0$ 
(e.g.\ \cite{BBL11}), 
by varying $\alpha$ we obtain a ``natural'' representative 
for each regularity index at $0$.
Finally, the Beta$(2-\alpha, \alpha)$-coalescents appear as limiting 
genealogies for a ``natural'' class of reproduction models 
derived from a branching mechanism with ``heavy tails'',  
see Section~\ref{subsec:exmodels} below.

\subsection{Population models with highly skewed offspring distributions 
leading to exc{\colzwo eptional} genealogies}

We follow a classical framework of population genetics and regard neutral, 
non-overlapping discrete-generation population
models with exchangeable offspring distribution, that is, {\em Cannings
models}, see \cite{C74}, \cite{C75}. Consider a (haploid) population of fixed size $N$, let $t \in \N$ denote
the $t$-th generation. Let ${\bf \nu}^{(t)}:=\big(\nu_1^{(t)}, \dots, \nu_N^{(t)}\big)$ 
denote the vector of offspring replacing the $N$ individuals in the $t$-th generation,
where $\nu_k^{(t)}$ is the number of children of individual $k$.
We assume that the random vectors ${\bf \nu}^{(t)}$, with $t \in \N$, are independent and identically 
distributed, and furthermore that, for fixed $t$, the random variables $\nu^{(t)}_1,\dots,\nu^{(t)}_N$ are 
exchangeable. We write $\nu_i = \nu^{(t)}_i$ when speaking about 
distributional properties in a single generation, where $t$ is irrelevant. 
Note that exchangeability and constant population size force that
the expected number of offspring of each individual $i$ is one, i.e.~$\E[\nu_i]=1$.
Let
\begin{equation}
\label{cn}
c_N:=\frac{\E[\nu_i(\nu_i-1)]}{N-1}
\end{equation}
(this is the probability that two randomly drawn individuals from the 
same generation are siblings).
Note that the numerator equals the variance of the offspring distribution, i.e.~$\sigma^2_N= \E[\nu_i(\nu_i-1)]$ and, 
of course, is independent of $i$ due to exchangeability. A famous result by Kingman \cite{K82} shows that 
if $\sigma_N^2 \to \sigma^2 \in (0, \infty)$ as $N \to \infty$ (and suitable 
higher moment bounds hold), then the genealogy
of a sample of size $n$ taken from the limiting population (time-changed by $1/c_N$) will be given by Kingman's $n$-coalescent.

In \cite{MS01}, extending Kingman's original result, M\"ohle and Sagitov provide 
the necessary and sufficient criteria so that a limiting Cannings population
model has a non-trivial genealogy given by a $\Lambda$-coalescent. In particular, 
if we consider the population model on
the time scale $1/c_N$, then the limiting genealogy will be 
governed by a $\Lambda$-coalescent iff 
\begin{itemize}
\item $c_N \to 0$, as $N \to \infty$,
\item for all $y \in (0,1)$ with $\Lambda(\{y\})=0$, we have
  \begin{equation}\label{neun} 
    \frac{N}{c_N} \mathbb{P}\{\nu_1 > Ny \} \longrightarrow \int_{(y,1]} \frac{1}{x^2} \, \Lambda(dx), \quad \mbox{as}\;\; N \to \infty, 
  \end{equation}
\item and finally, for $i \neq j$, $\mathbb{E}[\nu_1(\nu_1-1)\, \nu_2(\nu_2-1)]/(N^2 c_N) \longrightarrow 0$ as $N \to \infty$.
\end{itemize}
Condition (\ref{neun}) shows how the distribution of the individual 
offspring numbers determines the measure $\Lambda$ from (\ref{Lambdarates}) 
that governs the transition rates in the coalescent. Indeed, (\ref{neun}) can be interpreted as
\[
\mathbb{P}\{\tfrac{\nu_1}{N} \approx y \} = \frac{c_N}{N y^2} \Lambda(dy).
\]
It in particular 
forces the offspring distribution {\em to be {\colzwo ``highly''} skewed}, 
i.e.~there must be reproduction events
where the number of offspring is of the same order as the total population 
size, which implies that the variance  $\sigma^2_N$ of $\nu_1$ blows
up as $N$ gets large. Note that this condition also implies that these extreme reproductive events
need to happen on the time-scale $1/c_N$, hence cannot be too frequent 
(an effect that has been pointed out, informally, in the introduction and also already in \cite{A04}).

\subsection{Examples for models with extreme reproductive behaviour}
\label{subsec:exmodels}

Several classes of Cannings models have been considered
recently in the literature in this context. For example, in \cite{EW06}, Eldon and Wakeley discuss a model of size $N$ where, 
in each generation, exactly one uniformly chosen individual reproduces and becomes 
parent of a fixed (non-random) positive fraction of the population, leading to a measure with an atom in~(0,1].

If one is interested in a model with variable (random) sweepstakes sizes, one might consider  
Schweinsberg's (\cite{S03}) model, which leads naturally to
Beta-coalescent based genealogies and is
related to $\alpha$-stable branching processes for some $\alpha \in (1,2]$. 
Again, consider a haploid population of size $N$. Reproduction is
assumed to happen in two steps. First, each individual spawns (independently) $\tilde \nu_k^{(t)}$
offspring, according to a probability distribution with a power law
tail-behaviour, i.e.
\begin{equation}
\label{eq:taildecay}
\P\{\tilde \nu_k^{(t)}  \ge l \} \sim C l^{-\alpha}, \quad \alpha \in (1,2], C >0,
\end{equation}
so that the amount of potential offspring has infinite variance.
We denote the resulting vector of (potential) offspring by
$
{\bf \tilde \nu}^{(t)} = \big(\tilde \nu_1^{(t)}, \dots, \tilde \nu_N^{(t)}\big)$.
Note that the components do not (yet) sum up to $N$, but to some random $\tilde N$ typically much larger than $N$.  Hence, in a second step, $N$
individuals are chosen uniformly at random from the $\tilde N$
potential offspring particles, so that we obtain the new offspring
vector ${\bf \nu}^{(t)} = (\nu_1^{(t)}, \dots, \nu_N^{(t)})$
where $\nu_i^{(t)}$ denotes the number of individuals drawn from the
$\tilde \nu_i^{(t)}$ potential children of parent $k$.

This model has some resemblance of so-called ``type-III
survivorship'': high fertility leads to excessive amount of offspring,
corresponding to the first reproduction step, whereas high mortality
early in life is modeled in the second step.  

\subsection{Inference methods}
\label{inference}

Choosing a suitable limit of a Cannings population model (e.g.~one of the models above)
determines an explicit probabilistic mechanism for the ancestral process, i.e.~the underlying class of 
$\Lambda$-coalescents. Still, one  needs to calibrate 
the corresponding parameters to the observed DNA sample in questions, thus inferring ``evolutionary parameters'' like the mutation rate. 
We pursue a maximum-likelihood approach based on the full sample information. 
We assume that mutations occur according to the infinite-sites model, i.e., 
each new mutation hits a novel site. 

A general probabilistic mechanism of obtaining DNA samples from a coalescent tree 
-- which is our model for the gene genealogy of the ancestral limit process of our population models
under consideration -- is described in detail in \cite{BBSb09}.
Here, we provide a quick overview.
To obtain a sample of size $n$, 
first run an $n$-$\Lambda$-coalescent to obtain a {\em rooted}
coalescent tree.  On this rooted tree with $n$ leaves (numbered from
$1$ to $n$), place mutations along the branches at rate $r$
to obtain a gene genealogy (here, $r$ is the scaled mutation rate).
Then, label these mutations randomly: Given there are $s$ mutations
in total, attach uniformly at random the labels from $1, \dots, s$ to these mutations.  
An observed genetic {\em type} ${\bf x}$ is then given by the sequence of labels of mutations following
its path backwards from a leaf to the root. 
When there are $d$ different types, 
we enumerate 
them randomly, from $1, \dots, d$. Note that $d \leq s$, since each type has at least one unique mutation. We then let 
$
[{\bf t}, {\bf n}]=\big[({\bf x}_1, \dots, {\bf x}_d), (n_1, \dots, n_d)\big] 
$
denote the pair consisting of the observed unordered $d$-tuple of types ${\bf t} $ and their respective
multiplicities ${\bf n}$. Note that $[{\bf t}, {\bf n}]$ equivalently describes a tree, commonly referred to as {\colzwo a} \emph{genetree}.
We will denote the distribution of such a data set or tree, depending on $\Lambda$ and mutation rate $r$, by $\P_{\Lambda, r}$.

We may then compute the likelihood of observed data $[{\bf t}, {\bf n}]$ under the ``parameter'' $\vartheta = (\Lambda, r)$, 
i.e.~$\P_\vartheta ([{\bf t}, {\bf n}]) := \P_{\Lambda, r}([{\bf t}, {\bf n}])$ recursively,
conditioning on the last event in the coalescent history, see \cite[Section~1.3]{BBSb09}.
Such a recursion may, for small sample sizes with few mutations, be solved numerically. However, for more complex samples,
Monte-Carlo methods, for example Importance Sampling, need to be employed. Such methods are being discussed in detail in \cite{BBSb09} and implemented in the program \texttt{MetaGeneTree}.\footnote{Version~0.1.2, available from 
\url{http://metagenetree.sourceforge.net}}

\section{Analysis of DNA sequence data}
\label{datappl}


\subsection{{\colzwo Description of underlying datasets}}

The Pacific Oysters dataset {\colzwo presented in} \cite{BBB94} was obtained as the result of a
restriction-enzyme digest of mitochondrial DNA taken from 159 Pacific oysters
({\em Crassostrea gigas}) from British Columbia. This digest can (in an ad hoc fashion) be interpreted as sequence
information resulting in 49 segregating sites or positions, where an
enzyme either cuts or leaves the DNA-molecule intact, depending on the
allele present. This pattern was then manually edited to resolve violations
of the infinitely many sites model, resulting in the exclusion of four
samples and five sites, see Appendix~\ref{app_details} for details.
This dataset {\colzwo has also been} analysed in \cite{EW06}, {\colzwo where} the authors already 
pointed out the underlying genealogy might not be adequately modelled by Kingman's coalescent.

The second set of DNA sequence data was {\colzwo discussed in} \cite{A04}. {\colzwo There, \'Arnason} 
combined several datasets, published in other works, from a 250 bp region of the mitochondrial cytochrome
{\em b} gene of the Atlantic cod ({\em Gadus morhua}). In \cite{A04}, {\colzwo he} provided a {\colzwo discussion} of the whole {\colzwo combined} dataset {\colzwo which unfortunately} turned out to be too {\colzwo large} to be 
treated by our exact likelihood methods. For this reason we analysed the smaller {\colzwo component} datasets {\colzwo described} in \cite{AP96, APP98, APKS00, CM91, CSHW95, PC93, SA03} separately. 
As \'Arnason pointed out the{\colzwo se} samples stem from 
various {\colzwo geographic locations} throughout the Atlantic. 
{\colzwo In our analysis,} we cho{\colzwo o}se the most abundant type to 
represent the ancestral type, {\colzwo and we} also consider {\colzwo summing out} over all 
possible ancestral types.  {\colzwo Again, s}ome of the samples violated the assumptions of the
infinitely many sites model. To cope with this, we considered the combined dataset of 
\cite{A04} and introduced a consistent pattern of parallel mutations to resolve all violations (again, see Appendix~\ref{app_details} for details). 
This procedure lead to a dataset that {\colzwo corresponds to} the phylogenetic maximum parsimony network
from Figure~2 of \cite[p.~1875]{A04}. We then analysed the respective subsamples specified in the 
different publications. Table~\ref{DtestsOyster} and Table~\ref{DtestsCod} show some characteristics of the datasets, and we refer to Appendix~\ref{app_details} for a more detailed description.


\subsection{Rejection of {\colzwo the} ``Kingman hypothesis''?}

In several datasets, in particular the one discussed in \cite{BBB94},
standard tests reject the ``Kingman hypothesis'', indicating that the genealogies 
underlying the observed datasets might not be adequately described by a Kingman-coalescent.
In particular, we consider Tajima's
$D$ and Fu \& Li's $D$ in each case. Recall that for a sample of $n$
sequences, Tajima's $D$ (see \cite{T89}) is based on the normalized difference between the mean number of pairwise
differences $\Delta_n$ and the weighted number of segregating sites $S_n$.
In a neutral, Kingman-coalescent based scenario, $D$ should be approximately 0.
Small values of $D$ indicate shallow genealogies, large values
indicate long internal branches. Another standard test statistic is Fu \& Li's $D$ (see \cite{FL93}). Again,
the test statistic is based on a standardized variable which is the 
difference between the number of mutations on external branches $\eta_e$, and the number of mutations on internal branches
$\eta_i$, multiplied by a weighting factor. Values for approximate ``confidence intervals" (CIs)  for each $D$ can be
found in Table~2 of \cite{FL93}.

\begin{table}[h!]
{\small
\begin{center}
\begin{tabular}{|l|r|r||c|c|c|c||c|c|c|c|}
\hline
Ref. & $n$ & $d$ & Tajima's $D$ & $\Delta_n$ & $S_n$ & CI &  Fu \& Li's $D$ & $\eta_i$ & $\eta_e$ & CI \\
\hline
\hline
\cite{BBB94} & 155 & 40 & -2.65 * & 0.94 & 44 & [-1.76, 2.10] & -6.6 * & 14 & 30 & [-2.38,1.63]\\
\hline
\end{tabular}
\smallskip

\caption{Statistical tests reject Kingman hypothesis for Pacific Oyster data}
\label{DtestsOyster}
\end{center}
}
\end{table}

\begin{table}[h!]
{\small
\begin{center}
\begin{tabular}{|l|r|r||c|c|c|c||c|c|c|c|}
\hline
Ref. & $n$ & $d$ & Tajima's $D$ & $\Delta_n$ & $S_n$ & CI &  Fu \& Li's $D$ & $\eta_i$ & $\eta_e$ & CI \\
\hline
\hline
\cite{AP96} & 100 & 11 & -0.66 & 1.44 & 10 & [-1.78, 2.07] & -1.50 & 6 & 4 & [-2.36, 1.61]\\
\hline
\cite{APP98} & 109 & 11 & -0.89 & 1.23 & 10 & [-1.78, 2.07] & -1.54 & 6 & 4 & [-2.36, 1.61]\\
\hline
\cite{APKS00} & 78 & 12 & -1.45 & 1.04 & 11 & [-1.79, 2.06] & -1.83 & 6 & 5 & [-2.38, 1.59]\\
\hline
\cite{CM91} & 55 & 12 & -1.87 * & 0.84 & 11 & [-1.8, 2.05] & -2.24 & 5 & 6 & [-2.45, 1.57] \\
\hline
\cite{CSHW95} & 236 & 16 & -2.11 * & 0.39 & 15 & [-1.75, 2.11] & -2.24 & 9 & 6 & [-2.25, 1.65] \\
\hline
\cite{PC93} & 103 & 10 & -2.12 * & 0.46 & 12 & [-1.78, 2.07] & -3.07 * & 5 & 7 & [-2.36, 1.61] \\
\hline
\cite{SA03} & 74 & 15 & -1.28 & 1.59 & 14 & [-1.79, 2.06] & -2.88 * & 6 & 8 & [-2.38, 1.59] \\
\hline
\end{tabular}
\smallskip

\caption{Rejection of Kingman hypothesis for some Atlantic Cod datasets}
\label{DtestsCod}
\end{center}
}
\end{table}

Table~\ref{DtestsOyster} and Table~\ref{DtestsCod} show the observed values for each $D$ 
and the corresponding $95\%$ confidence intervals
for the Pacific Oyster and Atlantic Cod datasets, respectively.
Since both are always negative for all datasets, 
there is a consistent, sometimes rather weak, sometimes significant {\colzwo (marked by an asterisk)} {\colzwo indication of }
%
a ``shallow'' genealogy. 

\subsection{Likelihood analysis}

{\colzwo Figures ~\ref{likelihood-surfaces} and~\ref{unrooted_likelihood-surfaces} in Appendix~\ref{Sec:Appendix} contain the likelihood surfaces for our pair of parameters $(r, \alpha)$, that is, the
coalescent time mutation rate and the parameter of the underlying Beta-coalescent. Both the rooted and unrooted
tree cases are presented, where in the former case we assumed  
the most frequent type to be ancestral.}
{\colzwo The results were obtained with the tool \texttt{MetaGeneTree} using the methods introduced} in \cite{BBSb09} and \cite{BB08a}, see Section~\ref{inference}. The surfaces were calculated on a discrete grid and the position of the maximum of the surface reported.

Table~\ref{BBBdataset} shows the maximum likelihood estimate for the Pacific Oyster dataset. The surface was obtained on a discrete grid with spacing (0.2,0.05). For each gridpoint the likelihood was estimated by performing $10^8$ independent runs of importance sampling using the proposal distribution~\cite[Definition~2.11]{BBSb09}. This proved sufficient to get an estimated relative error around 0.02. 
Note that our maximum likelihood estimate $\hat\alpha = 1.2$ agrees well with a recently obtained estimate by~\cite{E11} of $\hat\alpha = 1.203$ for the same dataset using methods based on the site frequency spectrum. 

The {\colzwo column called} ``rooted'' in Table~\ref{ResultCodData} shows the maximum likelihood estimates for the Atlantic cod datasets. The grid-spacing was (0.1,0.05), and all datasets except \cite{CSHW95} could be analysed using the exact recursive formula. For the latter dataset we employed  importance sampling with proposal distribution~\cite[Definition~2.11]{BBSb09} using \emph{driving values} to estimate the likelihood on several gridpoints from a single run of the importance sampling, as detailed in~\cite[Appendix~A.3]{BBSb09}. The grid-spacing for the driving values was chosen as (0.2,0.1) and we again used $10^8$ independent runs. We calculated the likelihood for each true gridpoint whose euclidean distance is less then (0.4,0.2) of the respective driving value. After combining the results, this proved sufficient to estimate the likelihood for each true gridpoint with an estimated relative error of approximately 0.01.

In the column titled ``unrooted'' we present the arg-maxima of the likelihood surfaces obtained by summing the likelihoods of all different samples obtained by choosing a different type to be the ancestral one (thus the likelihood for an unrooted genetree), following the methods introduced in \cite[Section~2.1]{GT95}.

The maximum likelihood estimates for the tree-shape parameter $\alpha$ for both datasets range from 1.25 to 1.65. Recalling that $\alpha = 2$ corresponds to Kingman's coalescent, these results {\colzwo indicate consistently} that the data is better explained by a genealogy allowing for multiple mergers than by a Kingman-{\colzwo based} genealogy. We will {\colzwo briefly} discuss possible explanations for this evidence of shallow genealogies in the next section. Again our estimates agree with the estimates of $\hat\alpha \approx 1.55$ in~\cite{E11} for the full dataset.

Note that the datasets used here contain no \emph{apriori}-information about the ancestral type, so the likelihood of the unrooted trees should be used for estimation. However, as seen in~Figures ~\ref{likelihood-surfaces} and~\ref{unrooted_likelihood-surfaces}, the position of the maximum does not differ severely in both analyses. A closer inspection of the calculations reveals that the sum of probabilities of the different rooted trees is dominated by the probability of the tree with the most frequent type ancestral. Thus the root used in the ``rooted''-case appears to be the most plausible choice. The only exception is given by dataset~\cite{PC93}, where the sum is dominated by two summands, one of them being the tree with the root chosen due to abundance. The second tree, however, was not obviously set apart from the rest.

\begin{table}[h!]
{\small
\begin{center}
\begin{tabular}{|c|c|c|}
  \hline
  Ref. & Location & $(\hat{r},\hat{\alpha})$ \\
  \hline
  \hline 
  \cite{BBB94} & British Columbia & (1.2, 1.15) \\
  \hline 
\end{tabular}
\end{center}
\smallskip 
\caption{Estimate for $r$ and $\alpha$ for the \cite{BBB94} Pacific Oyster dataset, with the most abundant type considered ancestral.}
\label{BBBdataset}
}
\end{table}
\begin{table}[h]
\begin{center}
{\small
\begin{tabular}{|l|l||c|c|}
\hline
\multicolumn{2}{|c||}{} & \multicolumn{2}{c|}{$(\hat{r},\hat{\alpha})$} \\ 
\hline
Ref. & Location & rooted & unrooted \\
\hline\hline
\cite{AP96} & Norway & (0.7, 1.65) & (0.7, 1.65) \\
\hline
\cite{APP98} & Baltic/ trans. area & (0.6, 1.55) & (0.5, 1.45) \\ 
\hline
\cite{APKS00} & Greenland subsample & (0.7, 1.5) & (0.7, 1.5) \\ 
\hline
\cite{CM91} & Norway/ Newfoundland & (0.8, 1.4) & (0.7, 1.35) \\ 
\hline
\cite{CSHW95} & Newfoundland & (0.6, 1.5) & (0.6, 1.45) \\ 
\hline
\cite{PC93} & Newfoundland & (0.6, 1.4) & (0.6, 1.4) \\ 
\hline
\cite{SA03} & Faroe Islands & (0.7, 1.3) & (0.7, 1.3) \\
\hline
\end{tabular}
}
\end{center}
\smallskip

\caption{Maximum likelihood estimates for the Atlantic cod datasets, for the ``rooted'' genetrees (most abundant type ancestral) and for the ``unrooted'' genetrees (summing over all possibilities of choosing ancestral types).}
\label{ResultCodData}
\end{table}


\section{Discussion}
\label{Discussion}

\subsection{Possible biological causes for shallow genealogies}
\label{shallowdiscussion}

The presence of ``Hedgecock's reproduction sweepstakes'' is only one possible cause 
for violations of the Kingman framework produced by shallow genealogies. 
We will (non-exhaustively) address some effects that could account for the observed degree and pattern of variability here 
(following a discussion of \'Arnason in \cite[pp.~1882]{A04}). 
Apart from a recent population bottleneck for the Pacific Oyster data, the variability could be caused by 
{\em selection}, either acting directly on the observed part of the
genome or in the ``background'', the presence of {\em frequent selective sweeps} or 
{geographical subdivision} (resulting e.g. from glaciation events). 

The observed
 mutations were synonymous or functionally equivalent replacements
 \cite[pp.~1882]{A04}. Thus, \'Arnason argues against 
 direct selective effects as follows: Selection acting on RNA
 products etc.\ would be weak purifying selection, not positive
 selection required to explain the observed pattern; furthermore,
 it seems rather unconceivable ``that by selecting at random
 a 250-bp fragment of a 16-kb chromosome one finds several selected
 sites and even balanced polymorphisms due to selection by the
 cellular machinery.''

 Concerning indirect selection acting on the mitochondrial genome,
 \cite[p.~1883]{A04} writes: ``[T]here might be frequent selective
 sweeps of mitochondrial variation, which through linkage have
 brought haplotypes to high frequencies.'' Indeed, thinking e.g.\ 
 of \cite{DS04, DS05}, recurrent selective sweeps could be a mechanism 
 explaining multiple mergers in genealogies. 
 \cite{A04}'s answer is: ``This explanation can account for the data but
 the main difficulty is to explain why there would be so
 much adaptive evolution going on for mitochondrial activity in cod.'' 
 Here a comparison of the mitochondrial genome and/or its protein products 
 over several fish species might reveal that much is conserved, possibly arguing 
 against frequent [and recent] selective sweeps.

Regarding the possibility of {\em Population structure}, either resulting from the population
 splitting into various subgroups in different refuges during the ice
 age(s) or due to local adaptations (which are linked to, but not
 visible in the observed region of the genome), resulting in overall
 balancing selection, \cite[p.~1883]{A04} writes: ``The shallowness
 of the genealogy is evidence against these explanations.'' 
 He points to the divergence observed in \cite{P} to ``calibrate'' what shallowness means for the cod. 
 Furthermore, concerning local refuges, \'Arnason argues that under
 this hypothesis, due to physical distance, one would expect the Baltic cod to have very
 different type configurations from the North Atlantic cod, which is
 not the case. 

Finally, \'Arnason identifies a {\em ``sweepstake''-like} mechanism as the most plausible cause, and
as discussed above, population models with Beta-coalescent based genealogies are 
compatible with this explanation.
If one believes in such a mechanism, this has significant implications for the estimation of parameters and the real-time embedding of various quantities, 
see Section~\ref{sec_realtime}.

The overview article of Hedgecock and Pudovkin \cite{HP11} provides a further, more detailed discussion of the concept of ``Sweepstakes reproductive success'' (SRS) and
of evidence for its presence in marine populations, together with a thorough literature review. Hedgecock and Pudovkin conclude that the ``development of statistical tools to help decide between different coalescent models and to draw inferences about demographic and genetic parameters of interest'' are welcome.
In a similar spirit, \cite[p.~1883]{A04} writes 
``Studies of temporal variation are called for to test it and
better resolve the differences between historical and
contemporary factors influencing variance in offspring
number and effective population sizes.''
We will come back to these points asking for statistical studies in Section \ref{ssn:model} below.


\subsection{Age of the most recent common ancestor}
\label{sec_realtime}

Assuming that the Beta-coalescent and the estimated parameter values 
present a {\colzwo reasonable} approximation {\colzwo to the real population mechanisms under consideration}, 
we calibrate our models on the Atlantic cod data, using the method described in \cite[Section~6]{GT94} 
adapted to $\Lambda$-coalescents, see \cite[Appendix~A.4]{BBSb09}. The maximum likelihood estimates from Table~\ref{ResultCodData} 
can be used to estimate the time to the most recent common ancestor ($T_\text{MRCA}$)
in coalescent-time units conditioned on the observed data.
For comparison, we also estimated the time to the most recent common ancestor assuming that the Kingman coalescent is the appropriate model for the genealogy and using the corresponding estimate for the mutation rate at $\alpha=2$.
In both cases, we estimated the value of the cumulative distribution function on 
a discrete grid with spacing 0.1 ranging from 0.3 to 4.0 using $10^8$ independent runs of the Markov chain.
The two columns in the middle of Table~\ref{tab_common_ancestor_beta} and Table~\ref{tab_common_ancestor_kingman} show the approximation of the expected value based on the empirical distribution function, as well as the corresponding 95\%-credibility interval assuming the Beta-coalescent respectively Kingman prior for the genealogy. For this we interpolated the distribution function using cubic splines in Mathematica 7.0 \cite{W07} and reported the respective 0.025- and 0.975-quantiles. Independent replicates indicate that the variance due to the Monte Carlo method is negligible, data not shown.

\begin{table}
	\begin{center}
\begin{tabular}{|l||r|c||r|c|}
\hline
	& \multicolumn{2}{c||}{coal. time}	& \multicolumn{2}{c|}{real time (in kya)} \\
\hline
Ref.	& est. mean	& CI	& est. mean	& CI \\
\hline
\hline
\cite{AP96}	& 1.59	& [0.70, 3.07]	& 115.5	& [50.9, 222.8] \\
\hline
\cite{APP98}	& 1.82	& [0.82, 3.47]	& 113.1	& [51.0, 215.5] \\
\hline
\cite{APKS00}	& 1.60	& [0.68, 3.11]	& 116.0	& [49.0, 225.7] \\
\hline
\cite{CM91}	& 1.38	& [0.55, 2.76]	& 114.8	& [45.2, 229.0] \\
\hline
\cite{CSHW95}	& 1.52	& [0.55, 3.11]	& 94.7	& [34.5, 193.1] \\
\hline
\cite{PC93}	& 1.86	& [0.75, 3.66]	& 115.5	& [46.4, 227.6] \\
\hline
\cite{SA03}	& 1.72	& [0.77, 3.25]	& 124.6	& [55.6, 235.9] \\
\hline
\end{tabular}
\smallskip

	\caption{Estimates for $T_\text{MRCA}$ given the different datasets assuming the Beta-coalescent as the true underlying model. The respective estimated mean is given together with the corresponding credibility interval (CI), both in coalescent time units and embedded in real time, based on $(\hat{r},\hat{\alpha})$ from Table~\ref{ResultCodData}.}
	\label{tab_common_ancestor_beta}
	\end{center}
\end{table}

\begin{table}
	\begin{center}
\begin{tabular}{|l||r|c||r|c|}
\hline
	& \multicolumn{2}{c||}{coal. time}	& \multicolumn{2}{c|}{real time (in kya)} \\
\hline
Ref.	& est. mean	& CI	& est. mean 	& CI \\
\hline
\hline
\cite{AP96}	& 1.19	& [0.58, 2.22]	& 148.0	& [72.7, 276.3] \\
\hline
\cite{APP98}	& 1.29	& [0.63, 2.41]	& 147.1	& [71.6, 274.4] \\
\hline
\cite{APKS00}	& 1.05	& [0.51, 1.96]	& 151.7	& [74.3, 284.3] \\
\hline
\cite{CM91}	& 0.90	& [0.45, 1.67]	& 158.0	& [78.6, 293.4] \\
\hline
\cite{CSHW95}	& 0.86	& [0.43, 1.61]	& 151.8	& [75.5, 284.2] \\
\hline
\cite{PC93}	& 1.12	& [0.53, 2.12]	& 162.8	& [77.5, 307.6] \\
\hline
\cite{SA03}	& 0.95	& [0.49, 1.69]	& 186.7	& [97.4, 333.6] \\
\hline
\end{tabular}
\smallskip

	\caption{Estimates for $T_\text{MRCA}$ given the different datasets assuming Kingman's coalescent. The respective estimated mean is given together with the corresponding credibility interval (CI), both in coalescent time units and embedded in real time.}
	\label{tab_common_ancestor_kingman}
	\end{center}
\end{table}

\begin{figure}
	\begin{center}
		\psfrag{ylab}{\hspace{-.7cm}{\small cum. prob.}}
		\psfrag{xlab}{\hspace{-1cm}{\small time in years}}
		\psfrag{0.0}{\scriptsize{0.0}}
		\psfrag{0.2}{\scriptsize{0.2}}
		\psfrag{0.4}{\scriptsize{0.4}}
		\psfrag{0.6}{\scriptsize{0.6}}
		\psfrag{0.8}{\scriptsize{0.8}}
		\psfrag{1.0}{\scriptsize{1.0}}
		\psfrag{0e+00}[cB][cB]{\scriptsize{$0$}}
		\psfrag{0}[cB][cB]{\scriptsize{$0$}}
		\psfrag{50000}[cB][cB]{}
		\psfrag{1e+05}[cB][cB]{\scriptsize{$10^5$}}
		\psfrag{100000}[cB][cB]{\scriptsize{$10^5$}}
		\psfrag{150000}[cB][cB]{}
		\psfrag{2e+05}{\scriptsize{$2\cdot 10^5$}}
		\psfrag{200000}{\scriptsize{$2\cdot 10^5$}}
		\psfrag{250000}[cB][cB]{}
		\psfrag{3e+05}{\scriptsize{$3\cdot 10^5$}}
		\psfrag{300000}{\scriptsize{$3\cdot 10^5$}}
		\psfrag{4e+05}{\scriptsize{$4\cdot 10^5$}}
		\psfrag{A96axxxx}{\scriptsize \cite{AP96}}
		\psfrag{C95a}{\scriptsize \cite{CSHW95}}
		\psfrag{S03a}{\scriptsize \cite{SA03}}
		\psfrag{A96k}{\scriptsize \cite{AP96} ($\alpha=2$)}
		\psfrag{C95k}{\scriptsize \cite{CSHW95} ($\alpha=2$)}
		\psfrag{S03kxxxxxxxxxxx}{\scriptsize \cite{SA03} ($\alpha=2$)}
		\subfigure[]{
			\includegraphics[scale=.3]{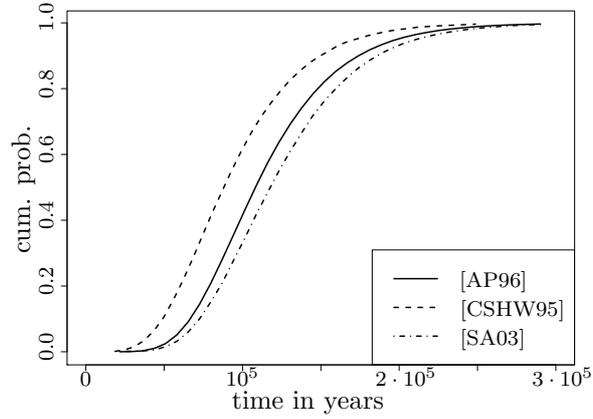}
			\label{inference_plot_common_ancestor_beta}
		}
		\subfigure[]{
			\includegraphics[scale=.3]{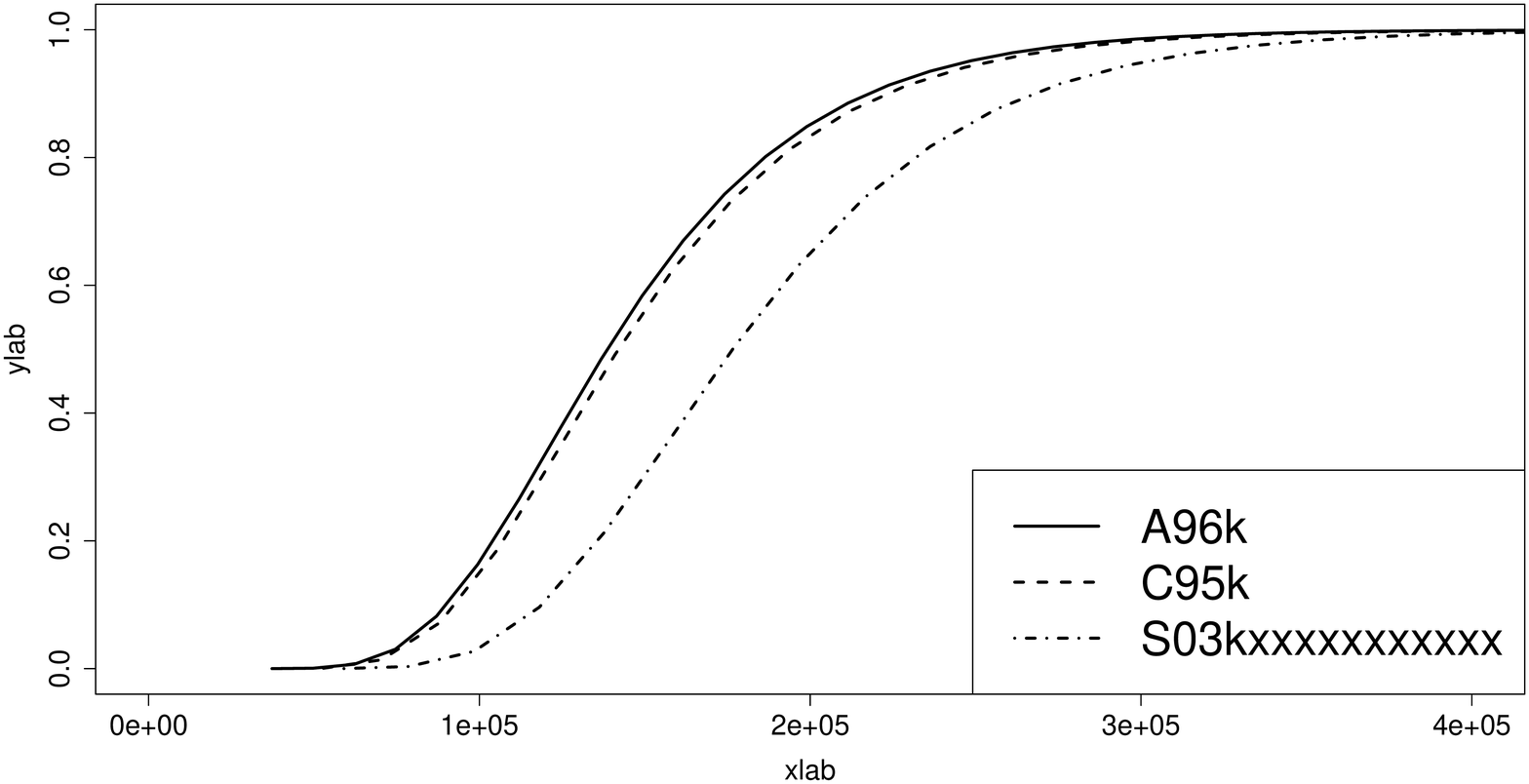}
			\label{inference_plot_common_ancestor_kingman}
		}
\caption{Distribution functions for the time to the most recent common 
ancestor of the sample, given the data, for three of the cod data sets, 
in the Beta- and Kingman-case. (a) The Beta-case, where we used the inferred parameter $\hat{\alpha}$ from Table~\ref{ResultCodData} as the underlying parameter. (b) The Kingman-case, where we used an underlying $\alpha=2$, as indicated in the legend.
}
		\label{inference_plot_common_ancestor}
	\end{center}
\end{figure}

To embed these values into real time {\colzwo  we use  \'Arnason's estimate (\cite[p.~1873]{A04}) for the mutation/substitution rate of the mitochondrial DNA for the Atlantic cod $\hat{\mu} = 3.86 \times 10^{-8}/\text{site}/\text{year}$}. 
Since we consider a stretch of 250 bp in the analysis of the Atlantic cod data, the substitution 
rate for this stretch is given by $\hat{\mu} = 9.65 \times 10^{-6}/\text{year}$. 
For the real time embedding of the coalescent time note that $\hat{\mu} \approx \#\text{mut.}/\text{year}$ and $\hat{r} \approx \#\text{mut.}/\text{coal.-time-unit}$. Thus the coalescent time can be transformed into real time by the relation
\[
\text{coal.-time-unit} \approx \frac{\text{year}}{\hat{r}}\hat{\mu}.
\]
The last two columns of Table~\ref{tab_common_ancestor_beta} and Table~\ref{tab_common_ancestor_kingman} show the real time embeddings (in kya) of the time to the most recent common ancestor and the corresponding credibility intervals for the different samples, in the Beta respectively Kingman case. Some cumulative distribution functions are shown in Figure~\ref{inference_plot_common_ancestor}.
We are not aware of any other estimates of $T_\text{MRCA}$ for the presented Atlantic cod datasets in the literature, but it would be interesting to compare our results with results obtained by other methods.

More importantly, our results show that the choice of the prior distribution for the genealogy has a severe impact on the estimation of $T_\text{MRCA}$. The estimates under the Kingman coalescent are approximately 50\% higher on average. This shows that, when estimating evolutionary parameters, it is crucial to verify the validity of Kingman's coalescent as an appropriate model for the underlying neutral genealogy.

\subsection{Further issues: Model-selection, allele-frequencies and statistical properties}
\label{ssn:model}

It is a natural question to ask, if one finds that Kingman's
coalescent is not a suitable approximation, e.g.\ due to the 
the presence of Hedgecock's ``sweepstakes reproductive success'' (SRS), 
which particular $\Lambda$-coalescents are of biological relevance. 
In order to do this, one needs to determine the distribution of the relative sizes of these reproduction sweepstakes to be won. This quantity typically depends in non-trivial ways on the offspring distribution of the species in question, but also on many other (ecological and demographic) factors which cannot be derived from simple assumptions from the outset (e.g.~the geography of the habitat).

In \cite{EW06}, Eldon and Wakeley derived and used (to fit to data 
from \cite{BBB94})
$\Lambda$-coalescents 
of the simple form $\Lambda=\delta_\psi$, thus positing a fixed 
relative sweepstake size $\psi \in (0,1]$, see 
Section~\ref{subset:exceptgenealog} for an interpretation. 
This class has the advantages of conceptual simplicity and that it depends 
only on one real parameter.
However, Eldon and Wakeley's model raises the question why SRS should always be a fixed fraction of size $\psi$ of the total population. As Hedgecock and Pudovkin point out: ``It should come as no surprise, then, that just as recruitment success in marine fisheries fluctuates greatly, so too may severity of SRS fluctuate'' (\cite[p.~973]{HP11}). 

In order to overcome this objection, while still keeping the statistical advantages of a class of coalescents parametrised by a single parameter, we chose the class of 
Beta$(2-\alpha, \alpha)$ coalescents. They arise naturally as scaling limits of branching-type populations models, where each individual reproduces independently (subject to preservation of the total population size) and offspring numbers are heavy-tailed with a power-law decay, see \eqref{eq:taildecay} and the subsequent discussion, in particular the relation to ``type-III-survivorship''.

Note that our scenario, motivated by Schweinsberg's model, considers
{\em infrequent} reproduction sweepstakes. This, together with a certain
amount of independence between reproductive events and subexponential tail distributions, makes it highly
unlikely to observe several ``sweepstakes winners'' within one generation
(hence ruling out $\Xi$-coalescent like behaviour): 
Typically, if a family of the order of the total population size is produced, 
the amount of offspring of the most successful individual dominates the number of offspring of the second-most successful
individual, 
thus ruling out, at least for large populations, the emergence of simultaneous multiple mergers. 


However, in other scenarios, other $\Lambda$-coalescents might become relevant. Examples include Durrett and Schweinsberg's models of recurrent selective sweeps \cite{DS05} mentioned above. 
The presence of externally induced recurrent severe bottlenecks might even yield $\Xi$-coalescent based genealogies, admitting {\em simultaneous} multiple merger coalescents \cite{BBMST09}.
Often, as in the bottleneck
scenario, simultaneous multiple mergers require drastic (externally
induced) changes in the environment, which we do not consider here.

Of course, stochastic fluctuations in the ocean environment (see \cite[p.~973]{HP11}),
could potentially cause such bottlenecks. It will be an objective of future research to test whether deviations from Kingman's coalescent seem to be induced predominantly by independent heavy-tailed reproduction, or by external stochastic fluctuations. However, such an analysis is clearly out of scope of this expository paper (in particular, there is a high risk of overfitting within the vast class of $\Xi$-coalescents).

Still, a way to assess the adequacy of the underlying coalescent models is to compare how well these models may be fitted to functions of the observed data such as frequency spectra. Indeed, in a recent presentation\footnote{Slides available at {\tt http://math.ucsd.edu/\~{}jschwein/LambdaSurvey.pdf}, see
Example 2.}, Schweinsberg 
compared minimal-least-squares fits of the observed {\em site frequency spectrum} of the full atlantic cod dataset from \cite{A04} (sample size $n=1278$) 
to the expectation in Beta$(2-\alpha, \alpha)$-models (using an asymptotically exact formula for the expected site frequency spectrum for the Beta$(2-\alpha, \alpha)$-case). 
The quality of the fit for the Beta-coalescent, with an estimated optimal $\alpha=1.43$, is striking 
and much better than for the Kingman-coalescent, making a strong point in favour of the use of 
Beta-coalescent models (the fit for the pacific oyster dataset from \cite{BBB94} mentioned in the same presentation 
is worse, however, as already mentioned above, in this case 
the population history suggests a recent severe bottleneck, which is 
not literally compatible with any of the coalescent models considered here).
\smallskip

It is a very interesting question in how far the DNA sample data
considered here allows a ``retrospective'' assessment of
``sweepstake sizes'' distributions, in particular whether a fixed
sweepstake size $\psi \in (0,1]$ as in the models considered by Eldon
and Wakely in \cite{EW06} appears more or less plausible than the 
``random'' one that is implicit in the Beta$(2-\alpha, \alpha)$-models. 

In preliminary computations, we computed likelihood values 
for the some of the component data sets considered here with $\Lambda$ as 
in (\ref{eq:EWatom}), varying $\psi$ on a grid in $(0,1]$. 
The maximal likelihood values, attained at $\psi$'s between $0.04$ and 
$0.07$, were sometimes comparable and sometimes one to two orders of 
magnitude smaller than those for the Beta$(2-\alpha, \alpha)$-coalescents 
(data not shown). In addition, we used the simulation algorithm described in 
\cite{BB08a} to estimate the expected site frequency spectrum under 
a coalescent with $\Lambda$ as in (\ref{eq:EWatom}) with $n=1278$; the 
fit to the observed frequency spectrum of the total sample described 
in \cite{A04} 
appeared much worse than that 
derived from the Beta$(2-\alpha, \alpha)$-coalescent described by 
Schweinsberg (data not shown).
While this suggests that ``non-fixed sweepstake sizes'' might be 
indeed a more reasonable model, it also indicates that larger sample sizes 
and presumably also multi-locus data sets would be required for a 
reliable answer. This is beyond the scope of the present work. 

Finally, it is still a largely open question to assess the statistical properties of the estimator employed here, which is based on methods described in \cite{BBSb09}.
Some considerations in this direction can be found in \cite{S09}.

\section*{Acknowledgement}

M.S.\ was supported in part by a DFG IRTG 1339 scholarship, by DFG-fellowship STE 2011/1-1,
and NIH grant R01-GM094402.
J.B.\ is supported in part by DFG grant BL 1105/3-1.
M.B.\ is supported in part by DFG grant BI 1058/2-1.

The authors would like to thank Jay Taylor for many useful discussions,
both on theoretical background as well as the handling of the datasets.

We would also like to thank two anonymous referees for their careful
reading and comments which helped to improve the presentation of the
manuscript.


\appendix

\section{Detailed description of the datasets}
\label{app_details}

We now describe the datasets analyzed in this paper in more detail. The datasets together with the scripts for handling the analysis are available from the authors upon request.

The input for our method has to be a valid genetree (cf.\ Section~\ref{inference}) under the infinitely-many-sites model, noting that DNA sequence data can only be transformed into this data structure if they satisfy the ``four-point'' condition~\cite[Equation~(10)]{BB08a}. We now provide details on how we dealt with occasional violations of this condition in the datasets that we analysed.

\subsection{Pacific Oyster dataset (\cite{BBB94})}

The dataset was taken from \cite{BBB94}, where the authors obtained the data as the result of a
restriction-enzyme digest of mitochondrial DNA taken from 155 Pacific oysters
({\em Crassostrea gigas}) from British Columbia. The authors reported the fragment sizes resulting from 9 different enzymes in their Table~2 \cite[pp.~1612-1613]{BBB94}.

We translated these lists of fragment length into pseudo-sequence data for every enzyme. Such a pseudo sequence is a list of zeros and ones for every reported type, specifying the presence or absence of a restriction enzyme binding site. The choice whether 1 denotes the presence of a binding site and 0 the absence or vice versa is determined uniquely once we specify an ancestral type later. Note that in~\cite[Table~2]{BBB94} the authors report a type `C' for the restriction enzyme `HincII', however, this type is not reported in \cite[Table~1]{BBB94}, so we omitted this type `C' for the subsequent analysis.

Table~1 of \cite[page 1610]{BBB94} shows how the different observed haplotypes are composed of the sub-haplotypes, along with the respective abundance of a given type in the respective sub-populations. Note that in~\cite[Table~1]{BBB94} type `HT32' reports a sub-type `F' for the enzyme `HAEII', but this sub-type `F' does not occur in the respective~\cite[Table~2]{BBB94}, thus we treated this sub-type as `E' in the subsequent analysis.

We chose the most abundant type ``PS2'' to be ancestral, thus uniquely specifying which pattern of presence/absence of restriction sites corresponds to the all-zero pseudo-sequence. To remedy the violations of the infinitely-many-sites model present in the dataset, we deleted the sites for `HindIII.5', `HaeIII.12', `AvaII.1', `HaeIII.11', `HaeII.6', and removed the types `HS45', `PS10', `HS44', `DB40'. Note that these steps eliminate the only difference between PS1 and PS2, so that ultimately PS1 and PS2 are taken to be ancestral. After these steps, the data can be converted into a valid genetree as described in Section~\ref{inference}.

\subsection{Atlantic Cod dataset(compiled in \cite{A04})}

The Atlantic Cod dataset is based on
DNA sequence data taken from a 250 bp stretch of the mitochondrial cytochrome
{\em b} gene of the Atlantic cod ({\em Gadus morhua}).  In the numbering of the
sites from \cite{JB96}, this stretch ranges from site 14,459 to site 14,708 (included). In \cite{A04}, \'Arnason took several cod datasets from different
publications, combined them and provided an analysis of the whole dataset.
However, since this combined dataset is too big to be treated by our method for
computing likelihoods based on the full information, we analysed certain samples separately. As \'Arnason pointed out, the samples stem from various
localities throughout the Atlantic, ranging from Newfoundland \cite{CM91},
\cite{PC93} \& \cite{CSHW95}, Greenland \cite{APKS00}, the Faroe Islands
\cite{SA03}, and Norway \cite{AP96} to the Baltic Sea \cite{APP98}.

The DNA sequence data of the different types present in the different samples
can be found in Figure~1 on page~1874 in \cite{A04}. The composition of the
sample from
\cite{AP96} is \{A: 35, E: 25, G: 14, D: 14, NI: 4, B: 2, C: 2, F: 1, GI: 1, DI: 1, BI: 1\}, the sample from
\cite{APP98} is \{E: 62, A: 19, G: 12, D: 6, DI: 3, H: 2, ES: 1, DK: 1, C: 1, EJ: 1, NI: 1\}, the sample from
\cite{APKS00} is \{A: 48, D: 6, E: 8, G: 7, C: 1, NI: 1, MI: 1, LI: 2, S: 1, PI: 1, GJ: 1, ED: 1\}, the sample from
\cite{CM91} is \{A: 36, B: 2, C: 2, D: 1, E: 4, F: 1, G: 4, H: 1, I: 1, J: 1, K: 1, L: 1\}, the sample from
\cite{CSHW95} is \{A: 201, B: 1, C: 2 , D: 7, E: 4, H: 3, J: 2, N: 4, O: 1, P: 1, S: 3, T: 1, X: 2, Y: 2, Z: 1, a: 1\}, the sample from
\cite{PC93} is \{A: 84, G: 6, E: 4, P: 1, X: 1, U: 2, N: 2, M: 1, C: 1, J: 1\}, and the sample from
\cite{SA03} is \{A: 26, E: 13, D: 11, G: 10, MI: 3, H: 2, NI: 1, C: 1, GJ: 1, EY: 1, EX: 1, EL: 1, EK: 1, DO: 1, DL: 1\}.

From \cite{APKS00} we took the Greenland subsample and we restricted the sample
from \cite{APP98} to the Baltic and transition area. In \cite{SA03}, the authors
provide the DNA sequence data of a larger 566 bp stretch from which we only took
the information of the 250 bp fragment in question. Furthermore, in \cite{CSHW95}, the authors report the type `R' which
differs from the type `A' only outside of the 250 bp segment we are considering.
Thus we count this type as an `A' type.

Since the full dataset from Figure~1 on page~1874 in \cite{A04} contains violations of the infinitely-many-sites model, we solved these violations by introducing a consistent pattern of parallel mutations. This procedure replaced each mutation violating the infinitely-many-sites model by a certain number of mutations that were attributed to the different types in a non-violating pattern. The complete
modified dataset with all parallel mutations is given in
Figure~\ref{arnason_consistent_dataset}. 

\begin{figure}
\renewcommand{\baselinestretch}{0.3}
\addtolength{\tabcolsep}{-5pt}
\begin{center}
{\tiny
\begin{tabular}{|l|cccccccccc|cccccccccc|cccccccccc|cccccccccc|cccccccccc|cccccccccc|r|}
\hline
\multicolumn{62}{|c|}{Segregating site}\\
\hline
       & 4& 4& 4& 4& 4& 4& 4& 4& 4& 4&  4& 4& 4& 5& 5& 5& 5& 5& 5& 5&  5& 5& 5& 5& 5& 5& 5& 5& 5& 5&  5& 5& 5& 6& 6& 6& 6& 6& 6& 6&  6& 6& 6& 6& 6& 6& 6& 6& 6& 6&  6& 6& 6& 6& 6& 6& 6& 6& 6& 7&\\
Haplo- & 6& 6& 6& 6& 6& 7& 8& 8& 8& 8&  9& 9& 9& 0& 0& 0& 1& 1& 2& 2&  3& 4& 4& 4& 4& 6& 6& 6& 6& 8&  8& 8& 9& 0& 3& 3& 3& 3& 3& 4&  4& 4& 5& 5& 5& 5& 6& 7& 7& 7&  8& 8& 9& 9& 9& 9& 9& 9& 9& 0&\\
type   & 3& 6& 7& 7& 8& 5& 1& 7& 7& 8&  0& 6& 6& 2& 8& 8& 4& 7& 2& 3&  5& 2& 4& 7& 7& 2& 2& 2& 5& 6&  6& 9& 5& 1& 1& 1& 1& 1& 7& 3&  9& 9& 8& 8& 8& 8& 4& 3& 6& 6&  5& 5& 1& 1& 1& 1& 5& 5& 7& 6&N\\
\hline
A   & .& .& .& .& .& .& .& .& .& .&  .& .& .& .& .& .& .& .& .& .&  .& .& .& .& .& .& .& .& .& .&  .& .& .& .& .& .& .& .& .& .&  .& .& .& .& .& .& .& .& .& .&  .& .& .& .& .& .& .& .& .& .& 696\\
E   & .& .& .& .& .& .& .& .& .& .&  .& .& .& .& p& .& .& .& .& .&  .& .& .& .& .& .& .& .& .& .&  .& .& .& .& .& .& .& .& .& .&  .& .& .& .& .& .& .& .& .& .&  .& .& .& .& .& .& .& .& .& .& 193\\
D   & .& .& .& .& .& .& .& .& .& .&  .& .& .& .& .& .& .& .& .& o&  .& .& .& .& .& .& .& .& .& .&  .& .& .& .& .& .& .& .& .& .&  .& .& .& .& .& .& .& .& .& .&  p& .& .& .& .& .& .& .& .& .& 124\\
G   & .& .& .& .& .& .& .& .& .& .&  .& .& .& .& .& .& .& .& .& .&  .& .& .& .& .& .& .& .& .& .&  .& .& .& .& .& .& .& .& .& .&  .& .& .& .& .& .& .& .& .& .&  .& .& .& .& .& p& .& .& .& .& 112\\
C   & .& .& .& .& .& .& .& .& .& .&  .& .& .& .& .& .& .& .& .& .&  .& .& .& .& .& .& .& .& .& .&  .& .& .& .& .& .& .& .& .& .&  .& .& .& .& .& .& .& .& .& .&  p& .& .& .& .& .& .& .& .& .&  29\\
NI  & .& .& .& .& .& .& .& .& .& .&  .& .& .& .& .& .& .& .& .& .&  .& .& .& .& .& .& .& .& .& .&  .& .& .& .& p& .& .& .& .& .&  .& .& .& .& .& .& .& .& .& .&  .& .& .& .& .& p& .& .& .& .&  15\\
H   & .& .& .& .& .& .& .& .& .& .&  .& .& .& .& .& .& .& .& .& .&  .& .& .& p& .& .& .& .& .& .&  .& .& .& .& .& .& .& .& .& .&  .& .& .& .& .& .& .& .& .& .&  .& .& .& .& .& .& .& .& .& .&   9\\
MI  & .& .& .& .& .& .& .& .& .& .&  .& .& .& .& .& .& .& .& .& .&  .& .& .& .& .& .& .& .& .& .&  .& .& .& .& .& p& .& .& .& .&  .& .& .& .& .& .& .& .& .& .&  .& .& .& .& .& .& .& .& .& .&   7\\
N   & .& .& .& .& .& .& .& .& .& o&  .& .& .& .& .& .& .& .& .& .&  .& .& .& .& .& .& .& .& .& .&  .& .& .& .& .& .& .& .& .& .&  .& .& .& .& .& .& .& .& .& .&  .& .& .& .& .& .& .& .& .& .&   6\\
J   & .& .& .& .& .& .& .& .& .& .&  .& .& .& .& .& .& .& .& .& o&  .& .& .& .& .& .& .& .& .& .&  .& .& .& .& .& .& .& .& .& o&  .& .& .& .& .& .& .& .& .& .&  p& .& .& .& .& .& .& .& .& .&   5\\
\hline
EC  & .& .& .& .& .& .& .& .& .& .&  .& .& .& .& p& .& .& .& .& .&  .& .& .& .& .& .& .& .& .& .&  .& .& .& .& .& .& .& .& .& .&  .& .& .& .& .& .& .& .& .& .&  .& p& .& .& .& .& .& .& .& .&   5\\
DI  & .& .& .& .& .& .& .& .& .& .&  .& .& .& .& .& .& .& .& .& o&  .& .& .& .& .& .& .& .& .& .&  .& .& .& .& .& .& .& .& .& .&  .& .& .& .& .& .& .& .& .& .&  p& .& p& .& .& .& .& .& .& .&   5\\
S   & .& .& .& .& .& .& .& .& .& .&  .& .& .& .& .& .& .& .& .& .&  .& .& .& .& .& .& .& .& .& .&  .& .& .& .& .& .& .& .& .& .&  p& .& .& .& .& .& .& .& .& .&  .& .& .& .& .& .& .& .& .& .&   4\\
AJ  & .& o& .& .& .& .& .& .& .& .&  .& .& .& .& .& .& .& .& .& .&  .& .& .& .& .& .& .& .& .& .&  .& .& .& .& .& .& .& .& .& .&  .& .& .& .& .& .& .& .& .& .&  .& .& .& .& .& .& .& .& .& .&   4\\
XI  & .& .& .& .& .& .& .& .& .& .&  o& .& .& .& p& .& .& .& .& .&  .& .& .& .& .& .& .& .& .& .&  .& .& .& .& .& .& .& .& .& .&  .& .& .& .& .& .& .& .& .& .&  .& .& .& .& .& .& .& .& .& .&   3\\
X   & .& .& .& .& .& .& .& .& .& .&  .& .& p& .& .& .& .& .& .& .&  .& .& .& .& .& .& .& .& .& .&  .& .& .& .& .& .& .& .& .& .&  .& .& .& .& .& .& .& .& .& .&  .& .& .& .& .& .& .& .& .& .&   3\\
RI  & .& .& .& .& .& .& .& .& .& .&  .& .& .& .& .& .& .& .& .& .&  .& .& .& .& .& .& .& .& .& .&  .& .& .& .& .& .& .& .& .& .&  .& .& .& .& .& .& .& .& .& .&  .& .& .& p& .& .& .& .& .& .&   3\\
LI  & .& .& .& .& .& .& .& .& p& .&  .& .& .& .& .& .& .& .& .& .&  .& .& .& .& .& .& .& .& .& .&  .& .& .& .& p& .& .& .& .& .&  .& .& .& .& .& .& .& .& .& .&  .& .& .& .& .& p& .& .& .& .&   3\\
BI  & .& .& .& .& .& .& .& .& .& .&  .& .& .& .& p& .& .& .& .& .&  .& .& .& .& .& p& .& .& .& .&  .& .& .& .& .& .& .& .& .& .&  .& .& .& .& .& .& .& .& .& .&  .& .& .& .& .& .& .& .& .& .&   3\\
B   & .& .& .& .& .& .& .& .& .& .&  .& .& .& .& .& .& .& .& .& .&  .& .& .& .& .& .& p& .& .& .&  .& .& .& .& .& .& .& .& .& .&  .& .& .& .& .& .& .& .& .& .&  .& .& .& .& .& .& .& .& .& .&   3\\
\hline
Y   & .& .& .& .& .& .& .& .& .& .&  .& .& .& .& .& .& .& .& .& .&  .& .& .& .& .& .& .& .& .& .&  .& .& .& .& .& .& .& .& .& .&  .& .& p& .& .& .& .& .& .& .&  .& .& .& .& .& .& .& .& .& .&   2\\
U   & .& .& p& .& .& .& .& .& .& .&  .& .& .& .& .& .& .& .& .& .&  .& .& .& .& .& .& .& .& .& .&  .& .& .& .& .& .& .& .& .& .&  .& .& .& .& .& .& .& .& .& .&  .& .& .& .& .& .& .& .& .& .&   2\\
P   & .& .& .& .& .& .& .& .& .& .&  .& .& .& .& .& .& .& .& .& .&  .& .& .& .& .& .& .& .& .& .&  .& .& .& .& .& .& .& .& o& .&  .& .& .& .& .& .& .& .& .& .&  .& .& .& .& .& .& .& .& .& .&   2\\
O   & .& .& .& .& .& .& .& .& .& .&  .& .& .& .& .& .& .& .& .& .&  .& .& .& .& .& .& .& .& .& .&  .& .& o& .& .& .& .& .& .& .&  .& .& .& .& .& .& .& .& .& .&  p& .& .& .& .& .& .& .& .& .&   2\\
LJ  & .& .& .& .& .& .& .& p& .& .&  .& .& .& .& .& .& .& .& .& .&  .& .& .& .& .& .& .& .& .& .&  .& .& .& .& .& .& .& .& .& .&  .& .& .& .& .& .& .& .& .& .&  .& .& .& .& .& p& .& .& .& .&   2\\
GJ  & .& .& .& .& .& .& .& .& .& .&  .& .& .& .& .& .& .& .& .& .&  .& .& .& .& .& .& .& .& .& .&  .& .& .& .& .& .& .& .& .& .&  .& .& .& .& .& .& .& .& .& p&  .& .& .& .& .& p& .& .& .& .&   2\\
AI  & .& .& .& .& .& .& .& .& .& .&  .& .& .& .& .& .& .& .& .& .&  .& .& .& .& .& .& .& .& .& .&  .& o& .& .& .& .& .& .& .& .&  .& .& .& .& .& .& .& .& .& .&  .& .& .& .& .& .& .& .& .& .&   2\\
a   & .& .& .& .& .& .& .& .& .& .&  .& .& .& .& .& .& .& o& .& .&  .& .& .& .& .& .& .& .& .& .&  .& .& .& .& .& .& .& .& .& .&  .& .& .& .& .& .& .& .& .& .&  .& .& .& .& .& .& .& .& .& .&   1\\
ZI  & .& .& .& .& o& .& .& .& .& .&  .& .& .& .& .& .& .& .& .& .&  .& .& .& .& .& .& .& .& .& .&  .& .& .& .& .& .& .& .& .& .&  .& .& .& .& .& .& .& .& .& .&  .& .& .& .& .& .& .& .& .& .&   1\\
Z   & .& .& .& .& .& .& .& .& .& .&  .& .& .& .& .& .& .& .& .& .&  .& .& .& .& .& .& .& .& .& .&  .& .& .& o& .& .& .& .& .& .&  .& .& .& .& .& .& .& .& .& .&  .& .& .& .& .& .& .& .& .& .&   1\\
\hline
UI  & .& .& .& p& .& .& .& .& .& .&  .& .& .& .& p& .& .& .& .& .&  .& .& .& .& .& .& .& .& .& .&  .& .& .& .& .& .& .& .& .& .&  .& .& .& .& .& .& .& .& .& .&  .& .& .& .& .& .& .& .& .& .&   1\\
TI  & .& .& .& .& .& .& .& .& .& .&  .& .& .& o& .& .& .& .& .& .&  .& .& .& .& .& .& .& .& .& .&  .& .& .& .& .& .& .& .& .& .&  .& .& .& .& .& .& .& .& .& .&  .& .& .& .& .& p& .& .& .& .&   1\\
T   & .& .& .& .& .& .& .& .& .& .&  .& .& .& .& .& .& .& .& .& .&  .& .& .& .& .& .& .& .& .& .&  .& .& .& .& .& .& .& .& .& .&  .& .& .& .& .& .& .& o& .& .&  .& .& .& .& .& .& .& .& .& .&   1\\
QI  & .& .& .& .& .& .& .& .& .& .&  .& .& .& .& .& .& .& .& .& .&  .& .& .& .& .& .& .& .& .& .&  .& .& .& .& .& .& .& .& .& .&  .& .& .& .& .& .& o& .& .& .&  .& .& .& .& .& .& .& .& .& .&   1\\
PI  & .& .& .& .& .& .& .& .& .& .&  .& .& .& .& .& .& .& .& .& .&  .& .& .& .& .& .& .& .& .& .&  .& .& .& .& .& .& .& .& .& .&  .& .& .& .& .& .& .& .& .& .&  .& .& .& .& .& .& p& .& .& .&   1\\
OI  & .& .& .& .& .& .& .& .& .& .&  .& .& .& .& p& .& .& .& .& .&  .& .& .& .& .& .& .& .& o& .&  .& .& .& .& .& .& .& .& .& .&  .& .& .& .& .& .& .& .& .& .&  .& .& .& .& .& .& .& .& .& .&   1\\
MC  & .& .& .& .& .& .& .& .& .& .&  .& .& .& .& .& .& .& .& .& .&  .& .& .& .& .& .& .& .& .& .&  .& .& .& .& .& .& .& .& .& .&  .& .& .& .& .& .& .& .& p& .&  p& .& .& .& .& .& .& p& .& .&   1\\
M   & o& .& .& .& .& .& .& .& .& .&  .& .& .& .& .& .& .& .& .& .&  .& .& .& .& .& .& .& .& .& .&  .& .& .& .& .& .& .& .& .& .&  .& .& .& p& .& .& .& .& .& .&  p& .& .& .& .& .& .& p& .& .&   1\\
L   & .& .& .& .& .& .& .& .& .& .&  .& .& .& .& .& .& .& .& .& .&  .& o& .& .& .& .& .& .& .& .&  .& .& .& .& .& .& .& .& .& .&  .& .& .& .& .& .& .& .& .& .&  .& .& .& .& .& .& .& .& .& .&   1\\
K   & .& .& .& .& .& .& .& .& .& .&  .& .& .& .& .& .& o& .& .& .&  .& .& .& .& .& .& .& .& .& .&  .& .& .& .& .& .& .& .& .& .&  .& .& .& .& .& .& .& .& .& .&  .& .& .& .& .& .& .& .& .& .&   1\\
\hline
I   & .& .& .& .& .& .& .& .& .& .&  .& .& .& .& .& .& .& .& .& .&  .& .& o& .& .& .& p& .& .& .&  .& .& .& .& .& .& .& .& .& .&  .& .& .& .& .& .& .& .& .& .&  .& .& .& .& .& .& .& .& .& .&   1\\
HJ  & .& .& .& .& .& .& .& .& .& .&  .& .& .& .& .& .& .& .& .& .&  .& .& .& p& .& .& .& .& .& .&  p& .& .& .& .& .& .& .& .& .&  .& .& .& .& .& .& .& .& .& .&  .& .& .& .& .& .& .& .& .& .&   1\\
HI  & .& .& .& .& .& .& .& .& .& .&  .& .& .& .& p& .& .& .& .& .&  .& .& .& .& p& .& .& .& .& .&  .& .& .& .& .& .& .& .& .& .&  .& .& .& .& .& .& .& .& .& .&  .& .& .& .& .& .& .& .& .& .&   1\\
GI  & .& .& .& .& .& .& .& .& .& .&  .& .& .& .& .& .& .& .& .& .&  .& .& .& .& .& .& .& .& .& p&  .& .& .& .& .& .& .& .& .& .&  .& .& .& .& .& .& .& .& .& .&  .& .& .& .& .& .& .& .& .& .&   1\\
F   & .& .& .& .& .& .& .& .& .& .&  .& .& .& .& p& .& .& .& o& .&  .& .& .& .& .& .& .& .& .& .&  .& .& .& .& .& .& .& .& .& .&  .& .& .& .& .& .& .& .& .& .&  .& .& .& .& .& .& .& .& .& .&   1\\
EY  & .& .& .& .& .& .& .& .& .& .&  .& .& .& .& p& .& .& .& .& .&  .& .& .& .& .& .& .& .& .& .&  .& .& .& .& .& .& .& .& .& .&  .& .& .& .& p& .& .& .& .& .&  .& .& .& .& .& .& .& .& .& .&   1\\
EX  & .& .& .& .& .& .& .& .& .& .&  .& p& .& .& p& .& .& .& .& .&  .& .& .& .& .& .& .& .& .& .&  .& .& .& .& .& .& .& .& .& .&  .& .& .& .& .& .& .& .& .& .&  .& .& .& .& .& .& .& .& .& .&   1\\
ES  & .& .& .& .& .& .& .& .& .& .&  .& .& .& .& p& .& .& .& .& .&  .& .& .& .& .& .& .& .& .& .&  .& .& .& .& .& .& .& .& .& .&  .& p& .& .& .& .& .& .& .& .&  .& .& .& .& .& .& .& .& .& .&   1\\
EM  & .& .& .& .& .& .& .& .& .& .&  .& .& .& .& p& .& .& .& .& .&  .& .& .& .& .& .& .& .& .& .&  .& .& .& .& .& .& .& p& .& .&  .& .& .& .& .& .& .& .& .& .&  .& .& .& .& .& .& .& .& .& .&   1\\
EL  & .& .& .& .& .& .& o& .& .& .&  .& .& .& .& p& .& .& .& .& .&  .& .& .& .& .& .& .& .& .& .&  .& .& .& .& .& .& .& .& .& .&  .& .& .& .& .& .& .& .& .& .&  .& .& .& .& .& .& .& .& .& .&   1\\
\hline
EK  & .& .& .& .& .& .& .& .& .& .&  .& .& .& .& p& .& .& .& .& .&  .& .& .& .& .& .& .& .& .& .&  .& .& .& .& .& .& .& .& .& .&  .& .& .& .& .& .& .& .& .& .&  .& .& .& .& .& .& .& .& o& .&   1\\
EJ  & .& .& .& .& .& .& .& .& .& .&  .& .& .& .& p& .& .& .& .& .&  .& .& .& .& .& .& .& .& .& .&  .& .& .& .& .& .& .& .& .& .&  .& .& .& .& .& .& .& .& .& .&  .& .& .& .& p& .& .& .& .& .&   1\\
EI  & .& .& .& .& .& o& .& .& .& .&  .& .& .& .& p& .& .& .& .& .&  .& .& .& .& .& .& .& .& .& .&  .& .& .& .& .& .& .& .& .& .&  .& .& .& .& .& .& .& .& .& .&  .& .& .& .& .& .& .& .& .& .&   1\\
ED  & .& .& .& .& .& .& .& .& .& .&  .& .& .& .& .& p& .& .& .& o&  .& .& .& .& .& .& .& .& .& .&  .& .& .& .& .& .& .& .& .& .&  .& .& .& .& .& .& .& .& .& .&  p& .& .& .& .& .& .& .& .& .&   1\\
DY  & .& .& .& .& .& .& .& .& .& .&  .& .& .& .& .& .& .& .& .& o&  .& .& .& .& .& .& .& .& .& .&  .& .& .& .& .& .& .& .& .& .&  .& .& .& .& .& p& .& .& .& .&  p& .& .& .& .& .& .& .& .& .&   1\\
DO  & .& .& .& .& .& .& .& .& .& .&  .& .& .& .& .& .& .& .& .& o&  o& .& .& .& .& .& .& .& .& .&  .& .& .& .& .& .& .& .& .& .&  .& .& .& .& .& .& .& .& .& .&  p& .& .& .& .& .& .& .& .& .&   1\\
DL  & .& .& .& .& .& .& .& .& .& .&  .& .& .& .& .& .& .& .& .& o&  .& .& .& .& .& .& .& .& .& .&  .& .& .& .& .& .& .& .& .& .&  .& .& .& .& .& .& .& .& .& .&  p& .& .& .& .& .& .& .& .& o&   1\\
DK  & .& .& .& .& .& .& .& .& .& .&  .& .& .& .& .& .& .& .& .& o&  .& .& .& .& .& .& .& .& .& .&  .& .& .& .& .& .& p& .& .& .&  .& .& .& .& .& .& .& .& .& .&  p& .& p& .& .& .& .& .& .& .&   1\\
DJ  & .& .& .& .& .& .& .& .& .& .&  .& .& .& .& .& .& .& .& .& o&  .& .& .& .& .& .& .& p& .& .&  .& .& .& .& .& .& .& .& .& .&  .& .& .& .& .& .& .& .& .& .&  p& .& .& .& .& .& .& .& .& .&   1\\
\hline
\end{tabular}
} 
\end{center}
\renewcommand{\baselinestretch}{1}
\caption{\'Arnason's dataset with violations of the IMS model resolved through parallel mutations. Here `o' denotes the original mutations (as in \'Arnason's dataset), whereas `p' denotes the putative parallel mutations.}
\label{arnason_consistent_dataset}
\end{figure}

We then obtained the subsamples corresponding to the different publications by choosing the
corresponding types in the corresponding quantities from this violation-free
dataset, and then converted them into genetrees. We chose `A' as the ancestral type for each dataset.

\section{Likelihood surfaces}
\label{Sec:Appendix}

The log${}_{10}$-likelihood surfaces for the cod and oyster datasets in the rooted case are shown in Figure~\ref{likelihood-surfaces}, whereas Figure~\ref{unrooted_likelihood-surfaces} shows the log${}_{10}$-likelihood surfaces for the cod datasets in the unrooted case.

\newpage

\setcounter{figure}{0}
\renewcommand{\thefigure}{\arabic{figure}}
\renewcommand{\thetable}{\arabic{table}}
\begin{figure}
        \centering
        \subfigure[\cite{AP96}]{
                \label{likelihood-surface1}
				\psfrag{main}{}
                \psfrag{xlab}{\small $\alpha$}
                \psfrag{ylab}{\small $r$}
                \includegraphics[width=40mm, height=40mm]{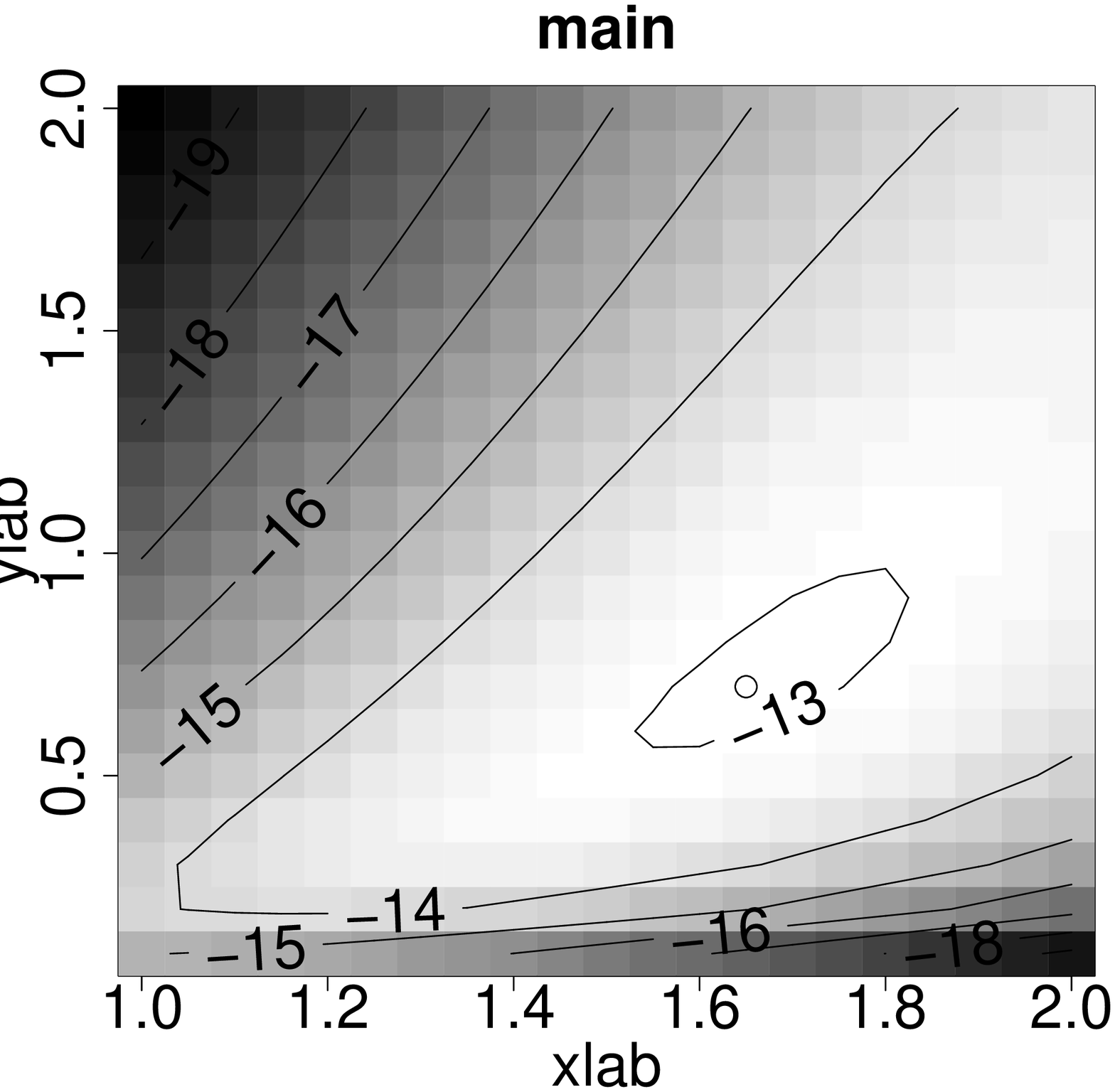}
        }
        \subfigure[\cite{APP98}]{
                \label{likelihood-surface2}
				\psfrag{main}{}
                \psfrag{xlab}{\small $\alpha$}
                \psfrag{ylab}{\small $r$}
                \includegraphics[width=40mm, height=40mm]{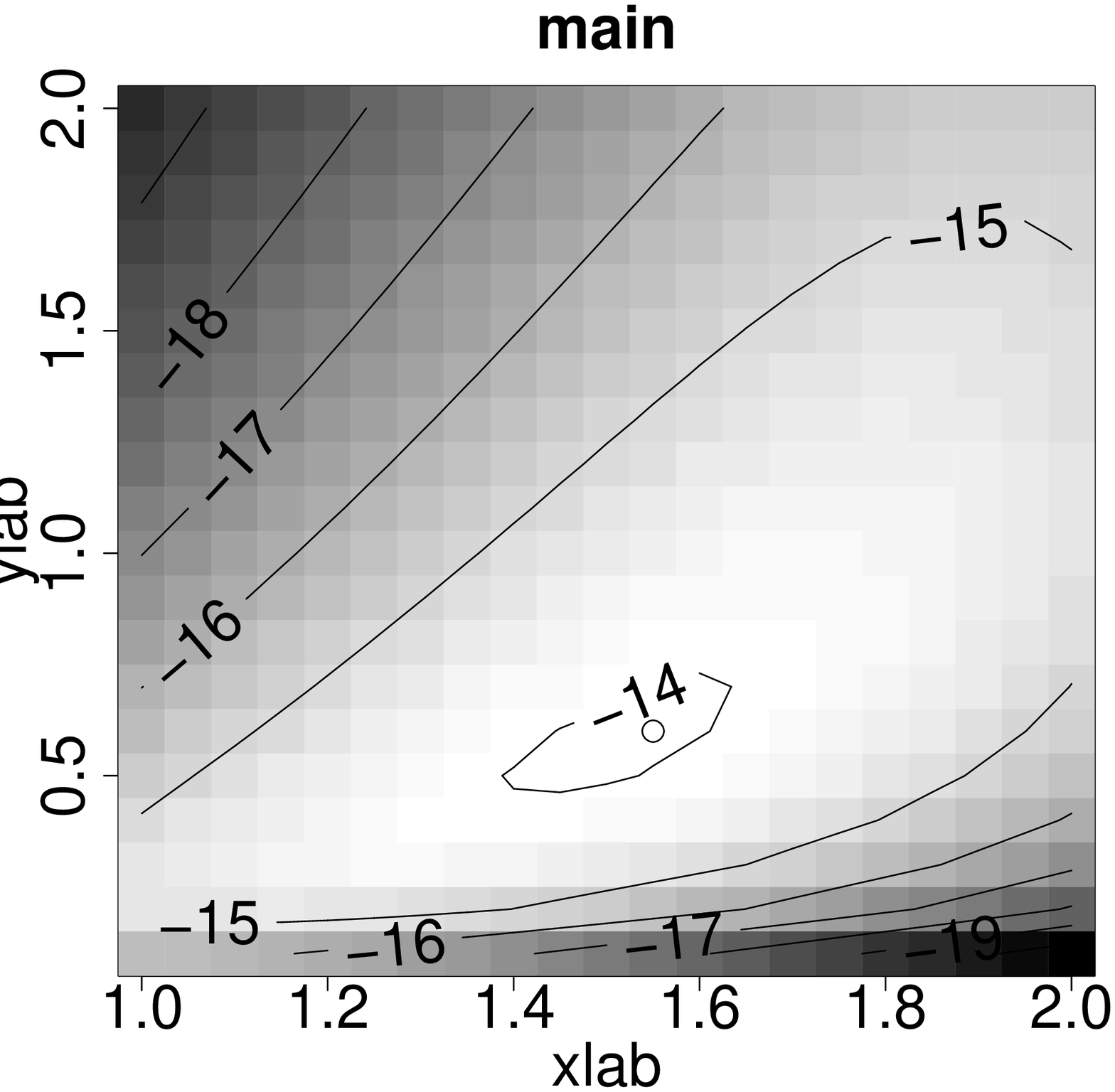}
        }
        \subfigure[\cite{APKS00}]{
                \label{likelihood-surface3}
				\psfrag{main}{}
                \psfrag{xlab}{\small $\alpha$}
                \psfrag{ylab}{\small $r$}
                \includegraphics[width=40mm, height=40mm]{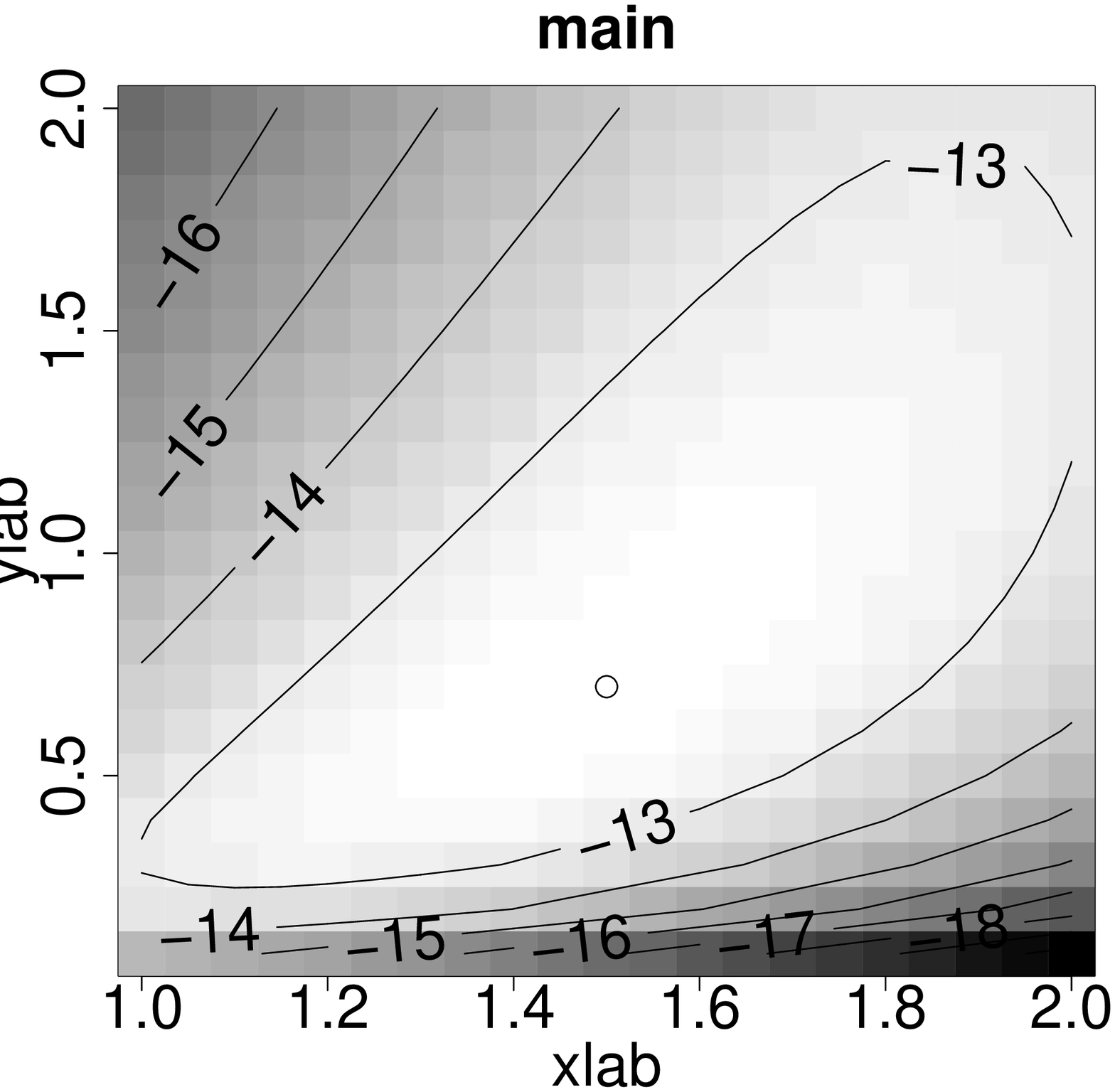}
        }\\[2ex]
        \subfigure[\cite{CM91}]{
                \label{likelihood-surface4}
				\psfrag{main}{}
                \psfrag{xlab}{\small $\alpha$}
                \psfrag{ylab}{\small $r$}
                \includegraphics[width=40mm, height=40mm]{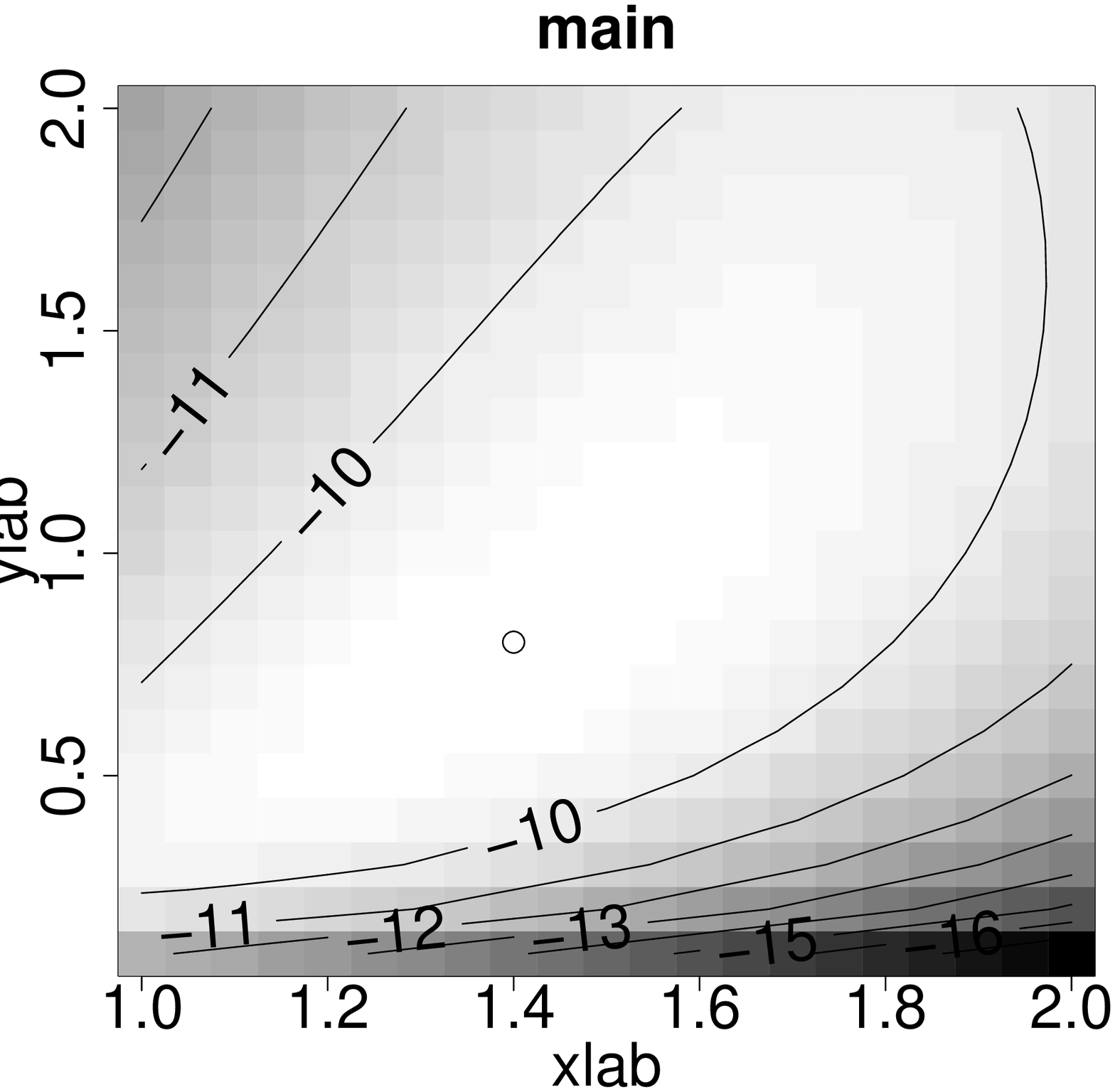}
        }
        \subfigure[\cite{CSHW95}]{
                \label{likelihood-surface5}
				\psfrag{main}{}
                \psfrag{xlab}{\small $\alpha$}
                \psfrag{ylab}{\small $r$}
                \includegraphics[width=40mm, height=40mm]{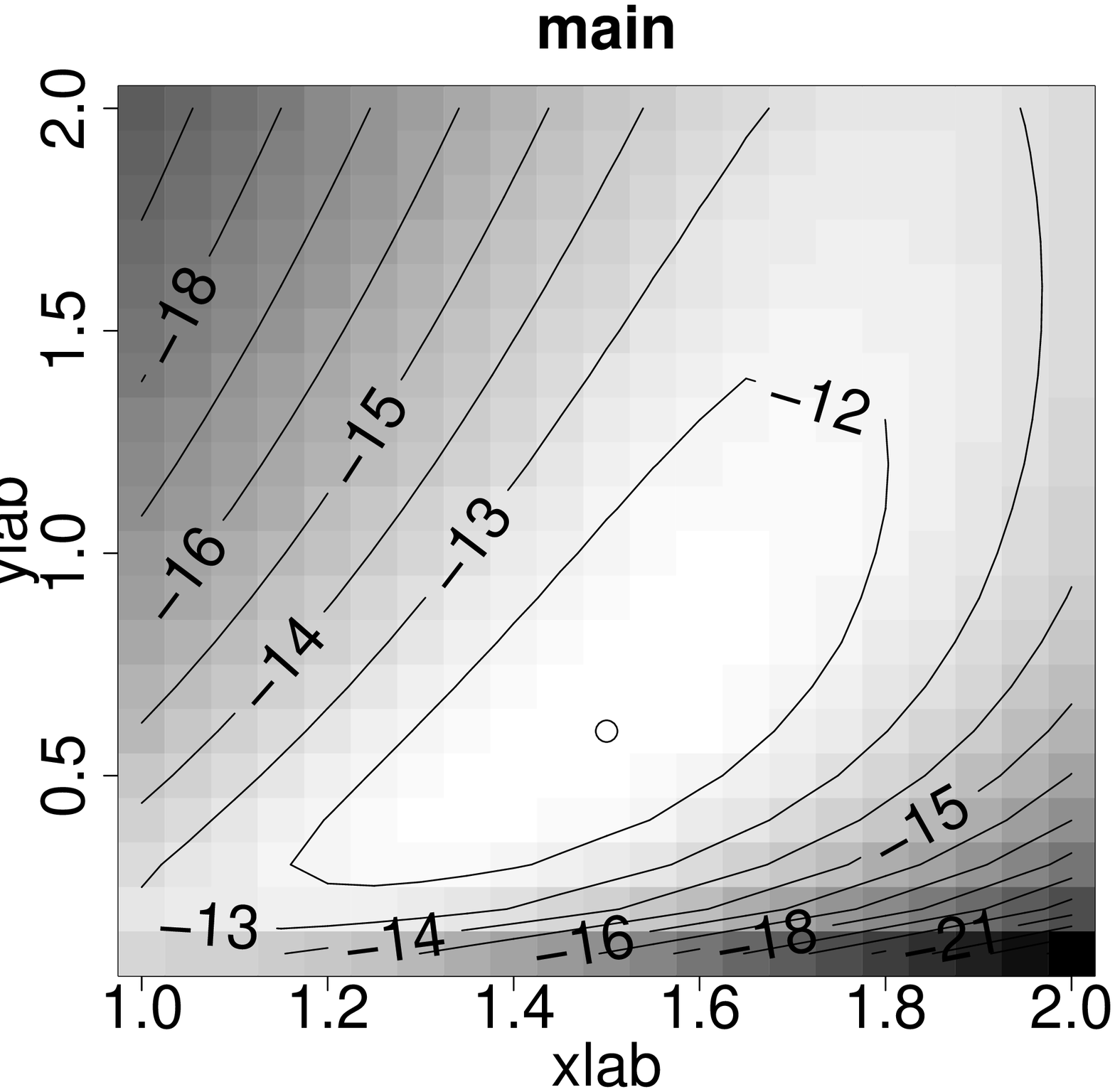}
        }
        \subfigure[\cite{PC93}]{
                \label{likelihood-surface6}
				\psfrag{main}{}
                \psfrag{xlab}{\small $\alpha$}
                \psfrag{ylab}{\small $r$}
                \includegraphics[width=40mm, height=40mm]{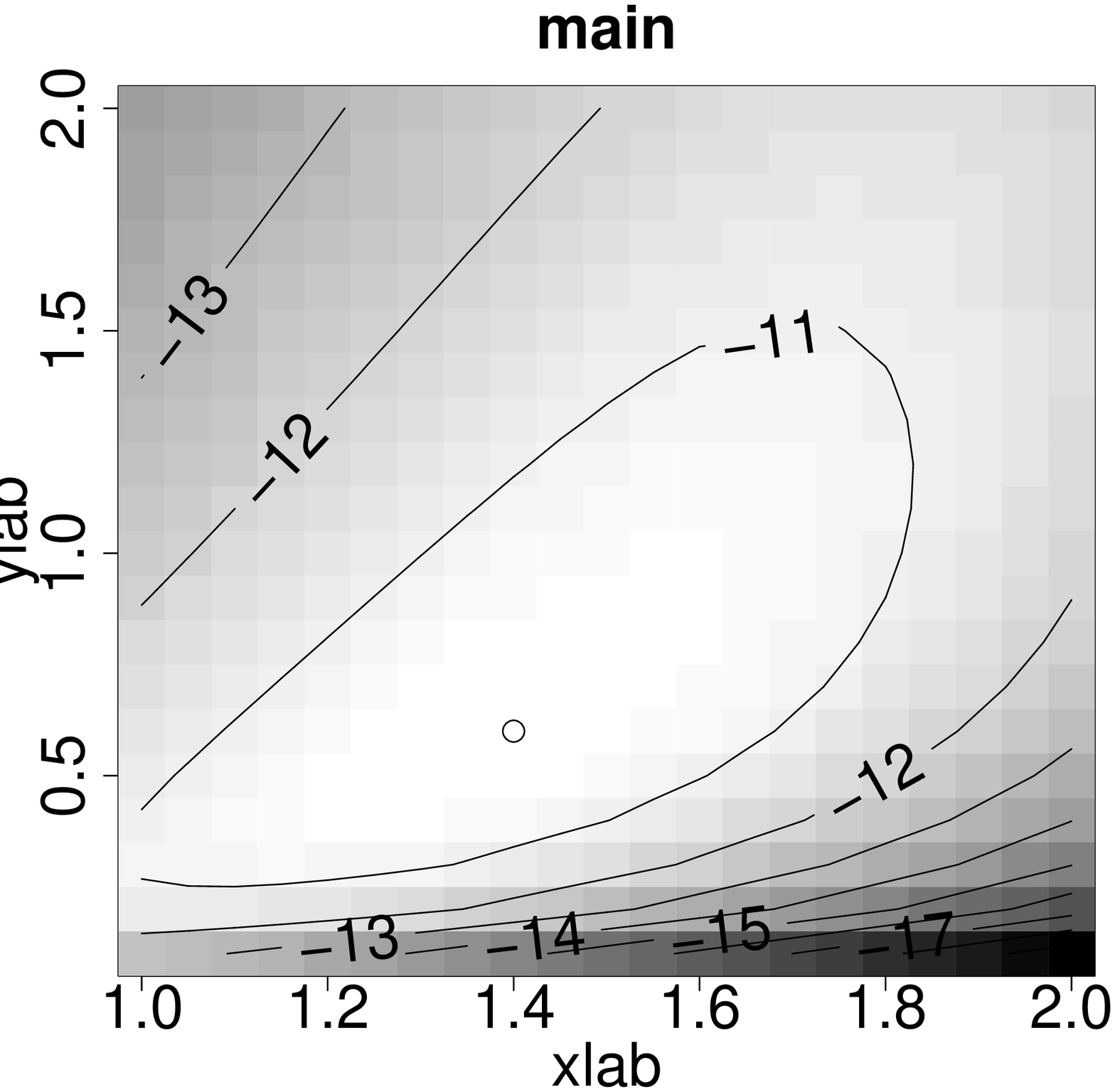}
        }\\[2ex]
        \subfigure[\cite{SA03}]{
                \label{likelihood-surface7}
				\psfrag{main}{}
                \psfrag{xlab}{\small $\alpha$}
                \psfrag{ylab}{\small $r$}
                \includegraphics[width=40mm, height=40mm]{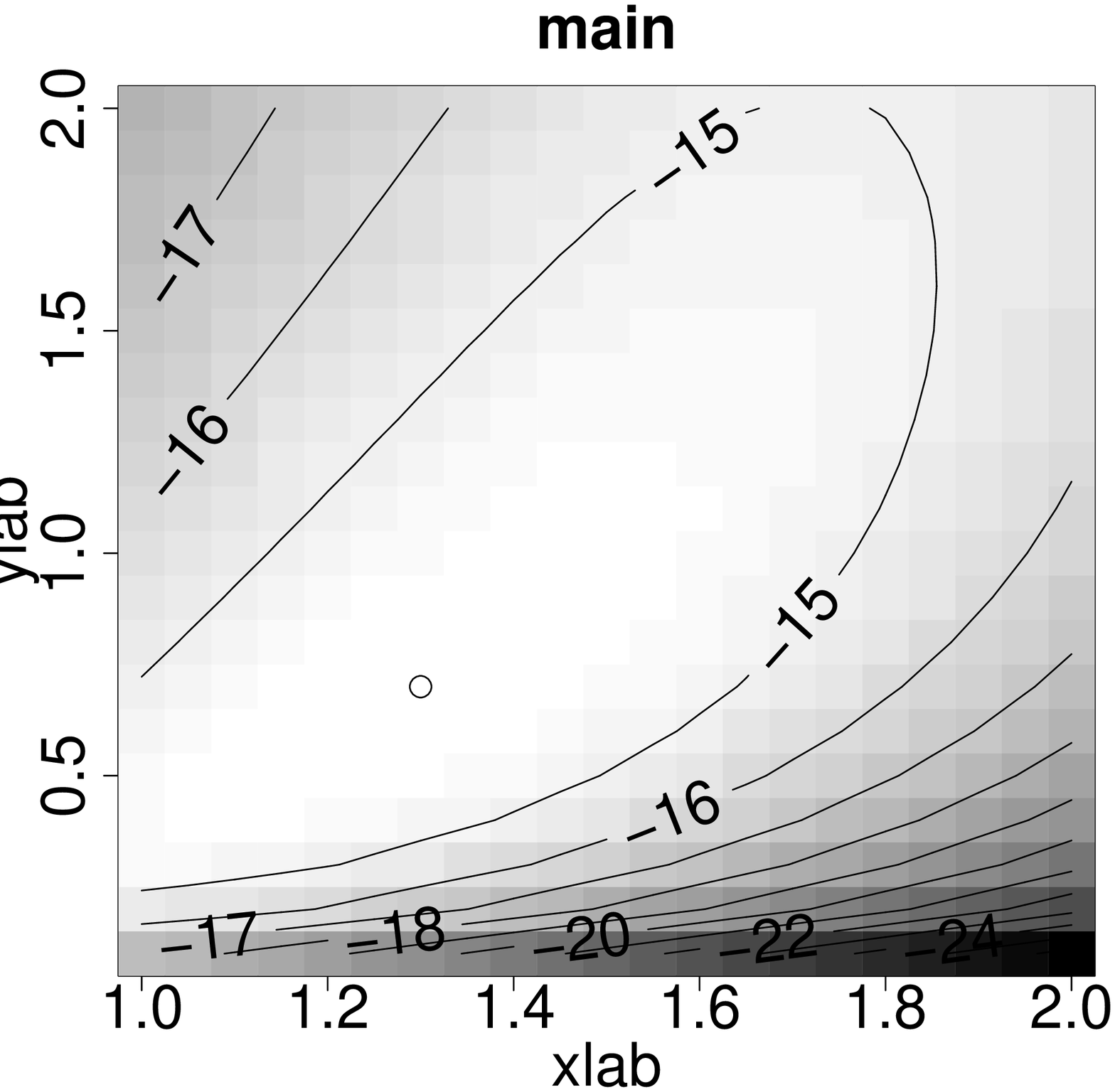}
                }
      \subfigure[(\cite{BB08a})]{ 
               \label{likelihood-surfaceOyster2}
				\psfrag{main}{}
                \psfrag{xlab}{\small $\alpha$}
                \psfrag{ylab}{\small $r$}
             	\includegraphics[width=40mm, height=40mm]{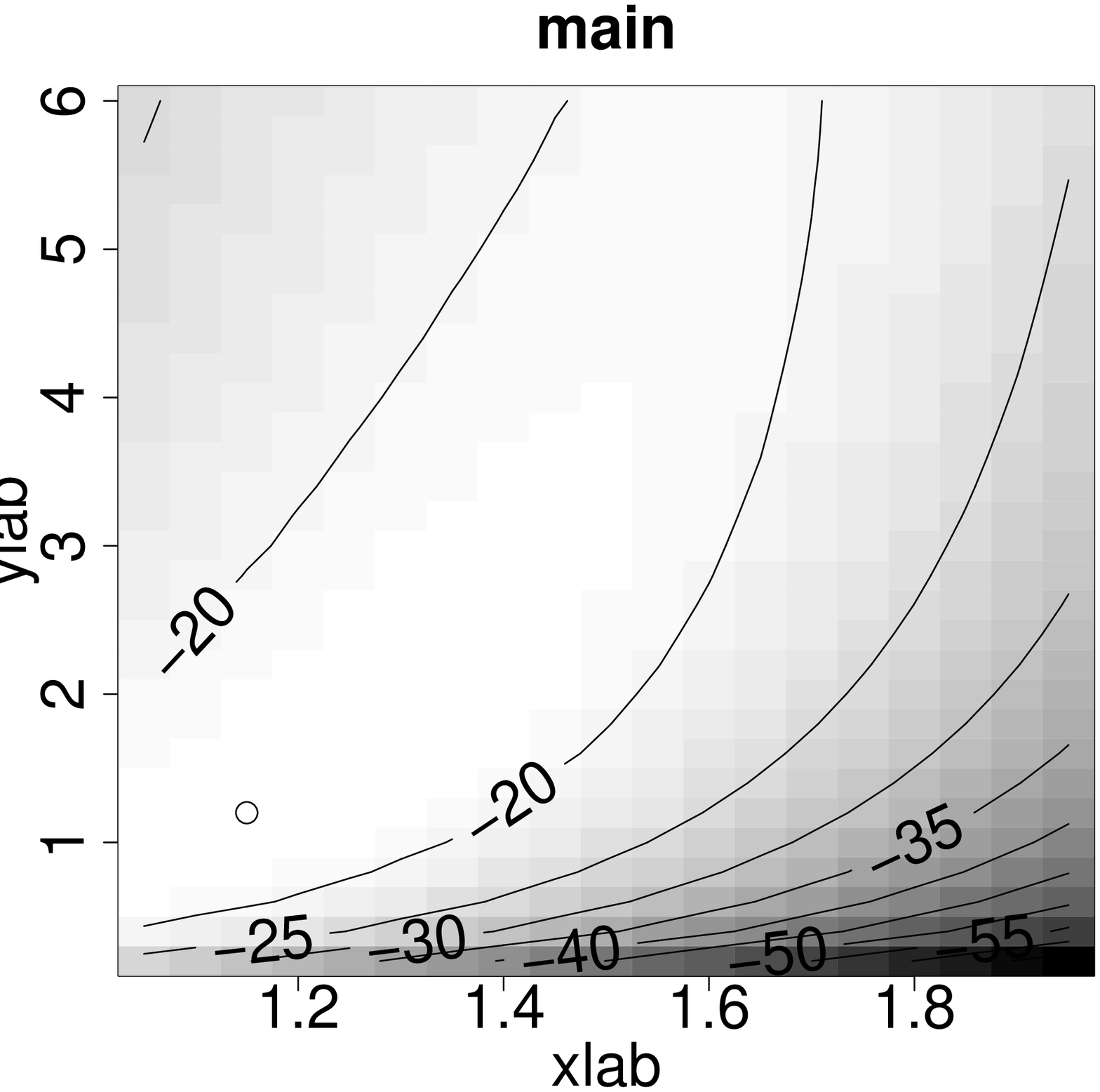}
        }
      \caption{Log${}_{10}$-Likelihood surfaces for cod and oyster datasets. 
The argmax is indicated by a dot.}\bigskip
      \label{likelihood-surfaces}
\end{figure}

\begin{figure}
        \centering
        \subfigure[\cite{AP96}]{
                \label{unrooted_likelihood-surface1}
				\psfrag{main}{}
                \psfrag{xlab}{\small $\alpha$}
                \psfrag{ylab}{\small $r$}
                \includegraphics[width=40mm, height=40mm]{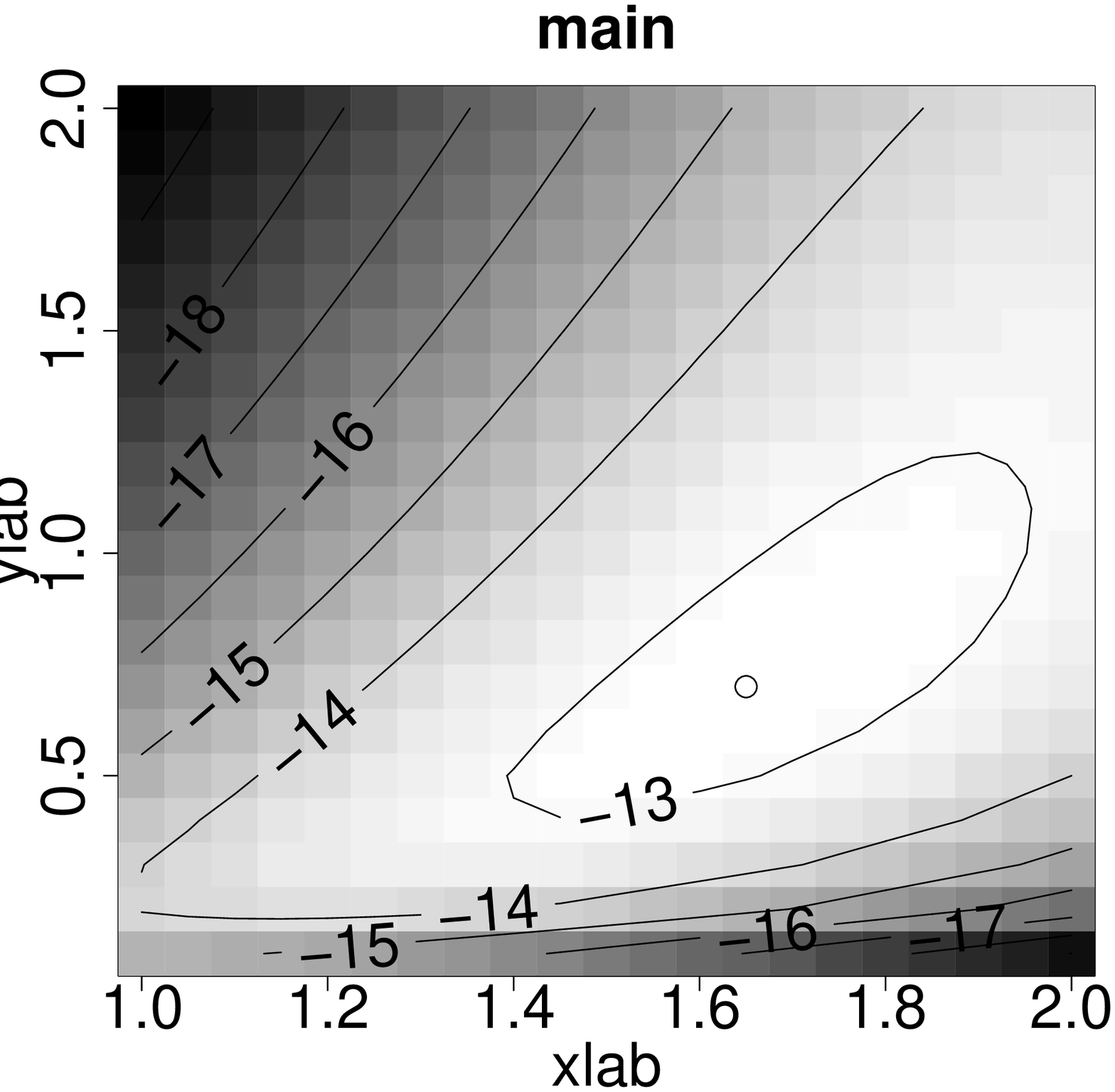}
        }
        \subfigure[\cite{APP98}]{
                \label{unrooted_likelihood-surface2}
				\psfrag{main}{}
                \psfrag{xlab}{\small $\alpha$}
                \psfrag{ylab}{\small $r$}
                \includegraphics[width=40mm, height=40mm]{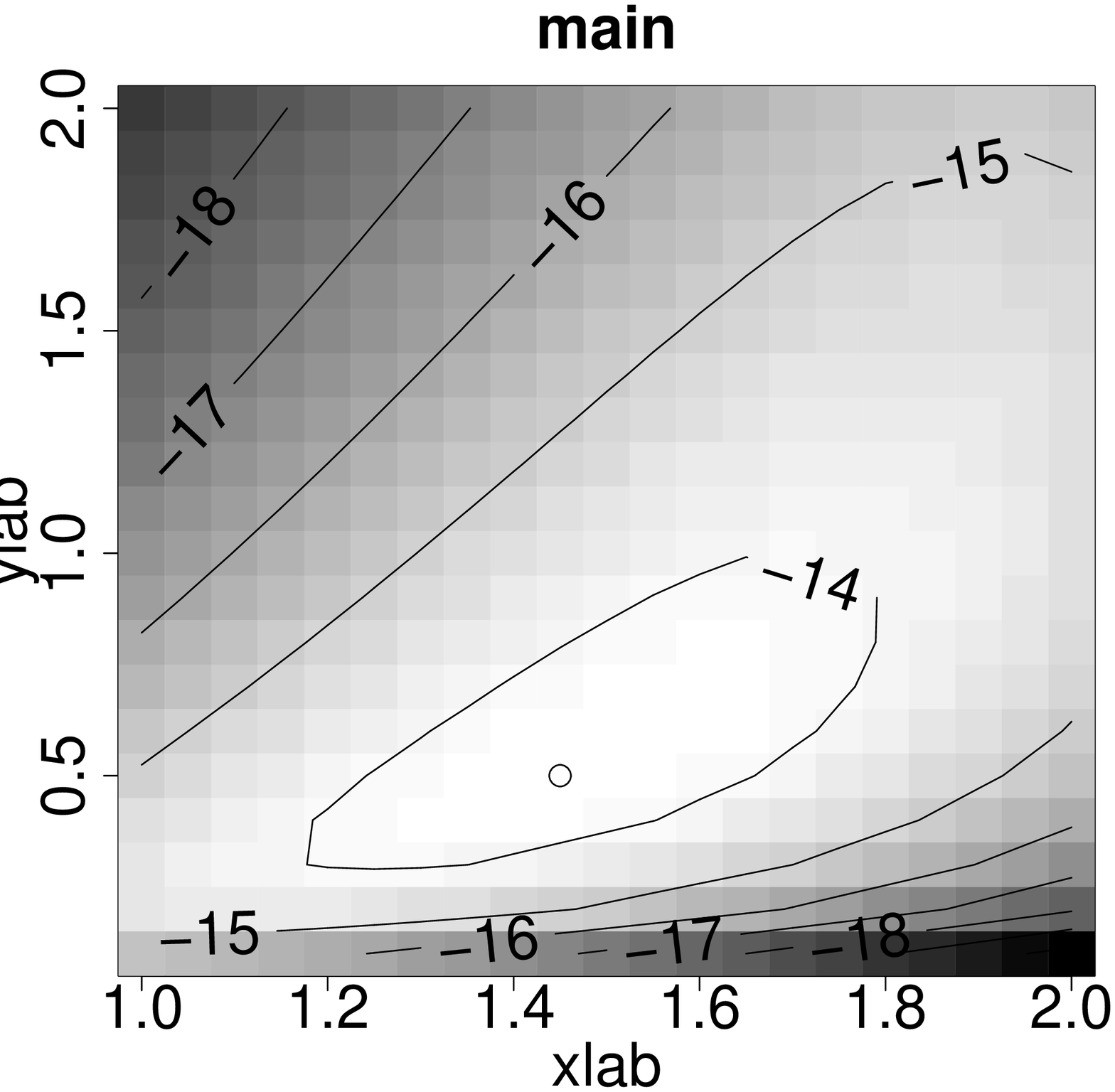}
        }
        \subfigure[\cite{APKS00}]{
                \label{unrooted_likelihood-surface3}
				\psfrag{main}{}
                \psfrag{xlab}{\small $\alpha$}
                \psfrag{ylab}{\small $r$}
                \includegraphics[width=40mm, height=40mm]{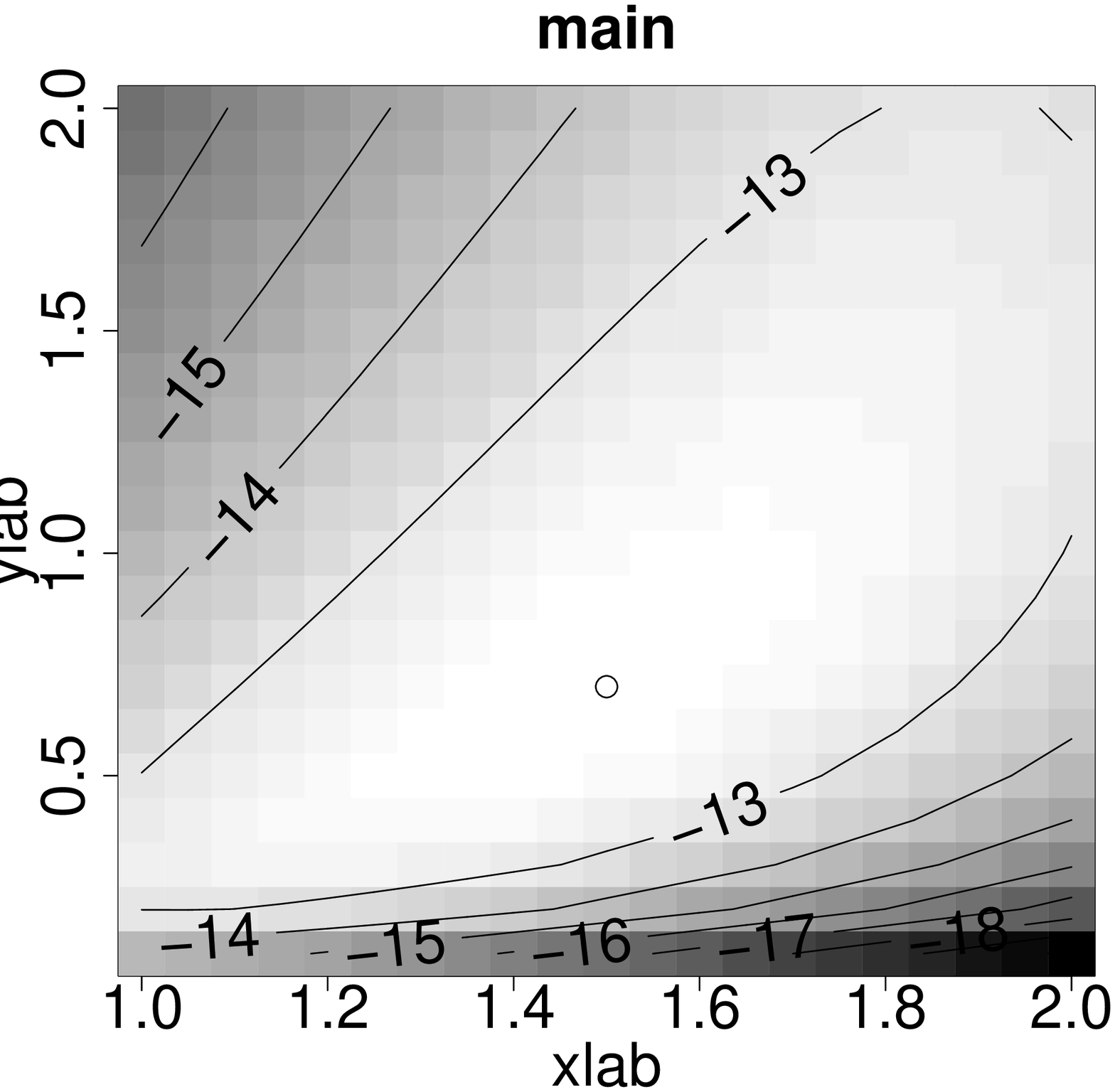}
        }\\[2ex]
        \subfigure[\cite{CM91}]{
                \label{unrooted_likelihood-surface4}
				\psfrag{main}{}
                \psfrag{xlab}{\small $\alpha$}
                \psfrag{ylab}{\small $r$}
                \includegraphics[width=40mm, height=40mm]{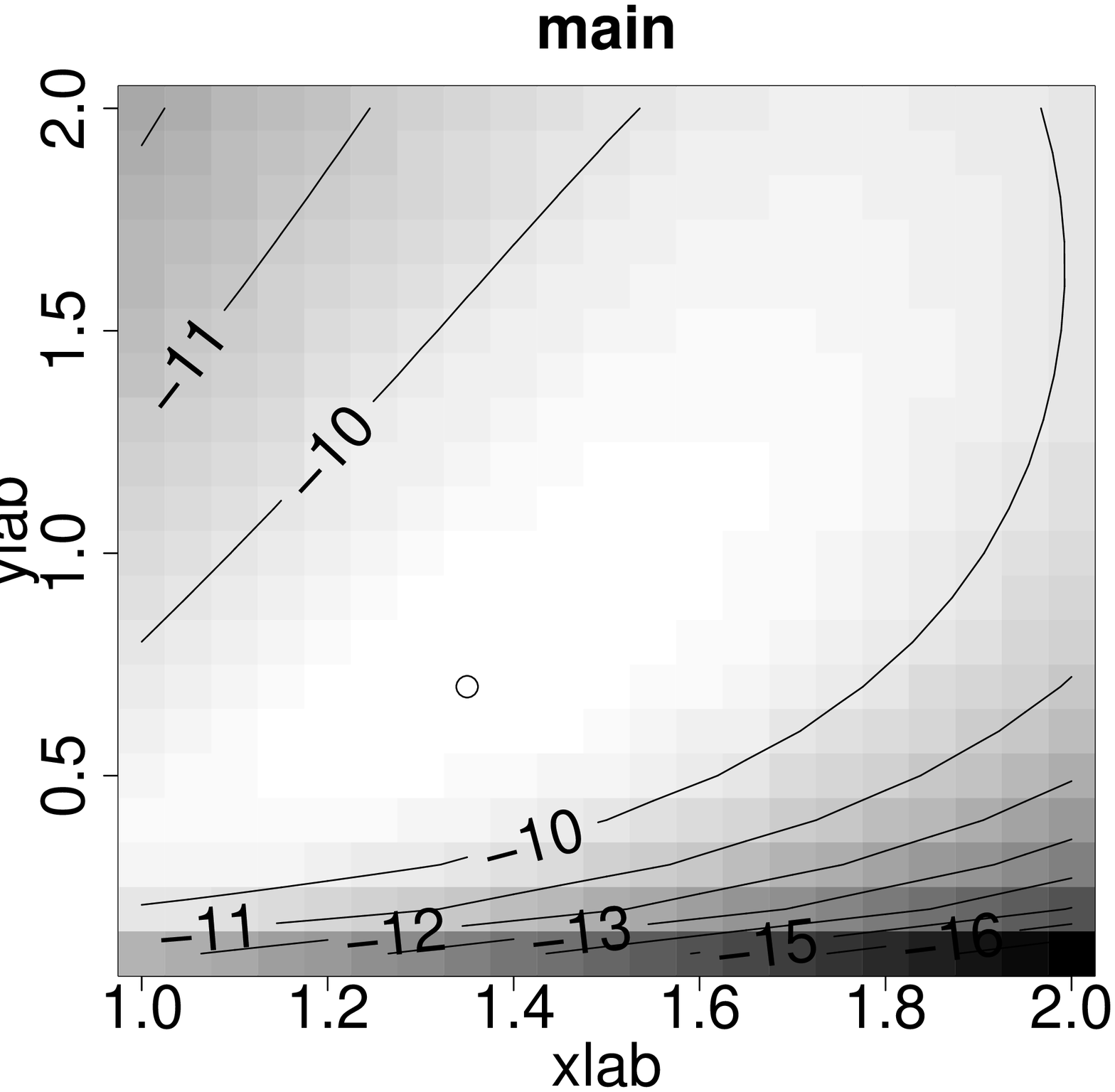}
        }
        \subfigure[\cite{CSHW95}]{
                \label{unrooted_likelihood-surface5}
				\psfrag{main}{}
                \psfrag{xlab}{\small $\alpha$}
                \psfrag{ylab}{\small $r$}
                \includegraphics[width=40mm, height=40mm]{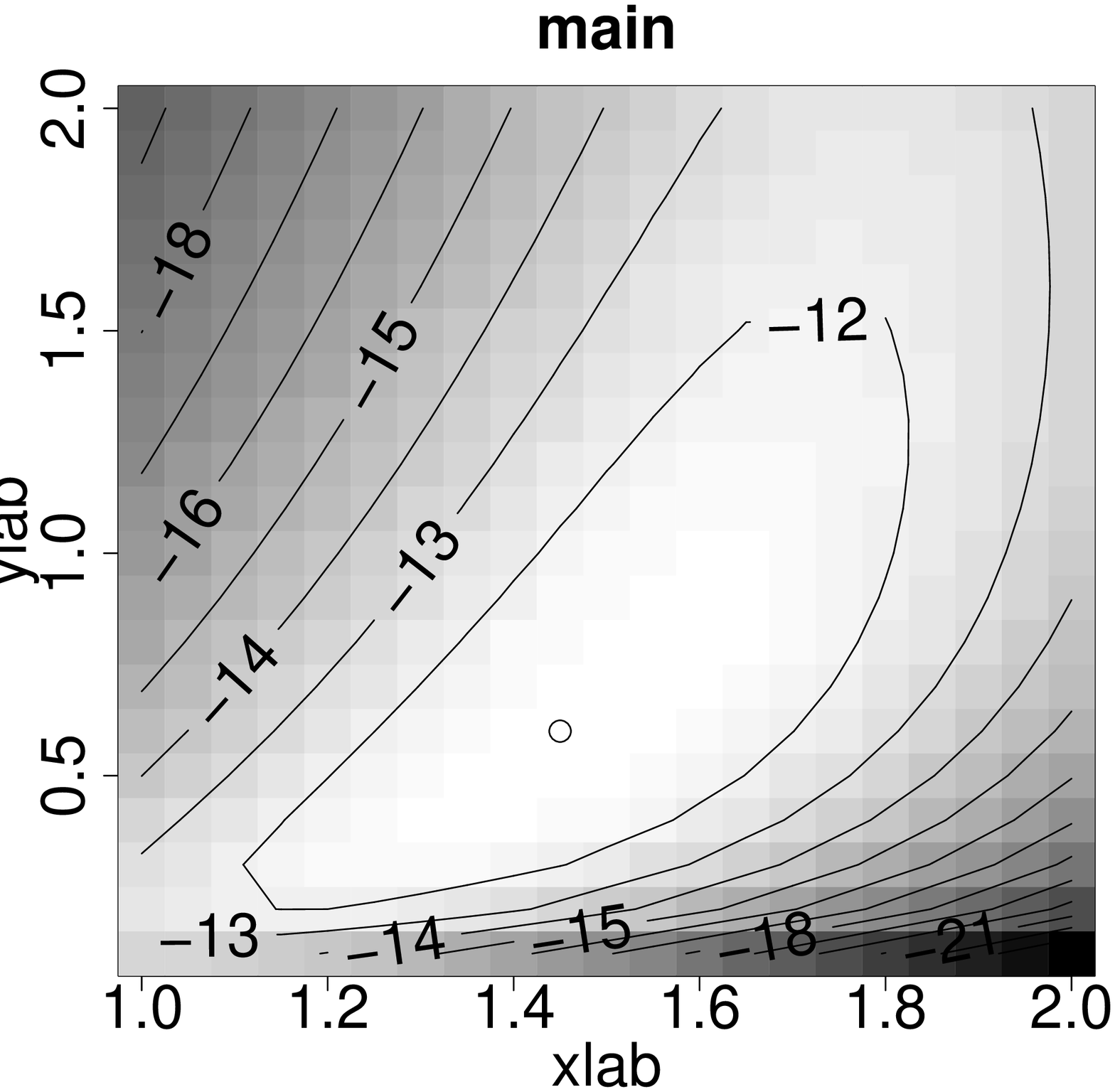}
        }
        \subfigure[\cite{PC93}]{
                \label{unrooted_likelihood-surface6}
				\psfrag{main}{}
                \psfrag{xlab}{\small $\alpha$}
                \psfrag{ylab}{\small $r$}
                \includegraphics[width=40mm, height=40mm]{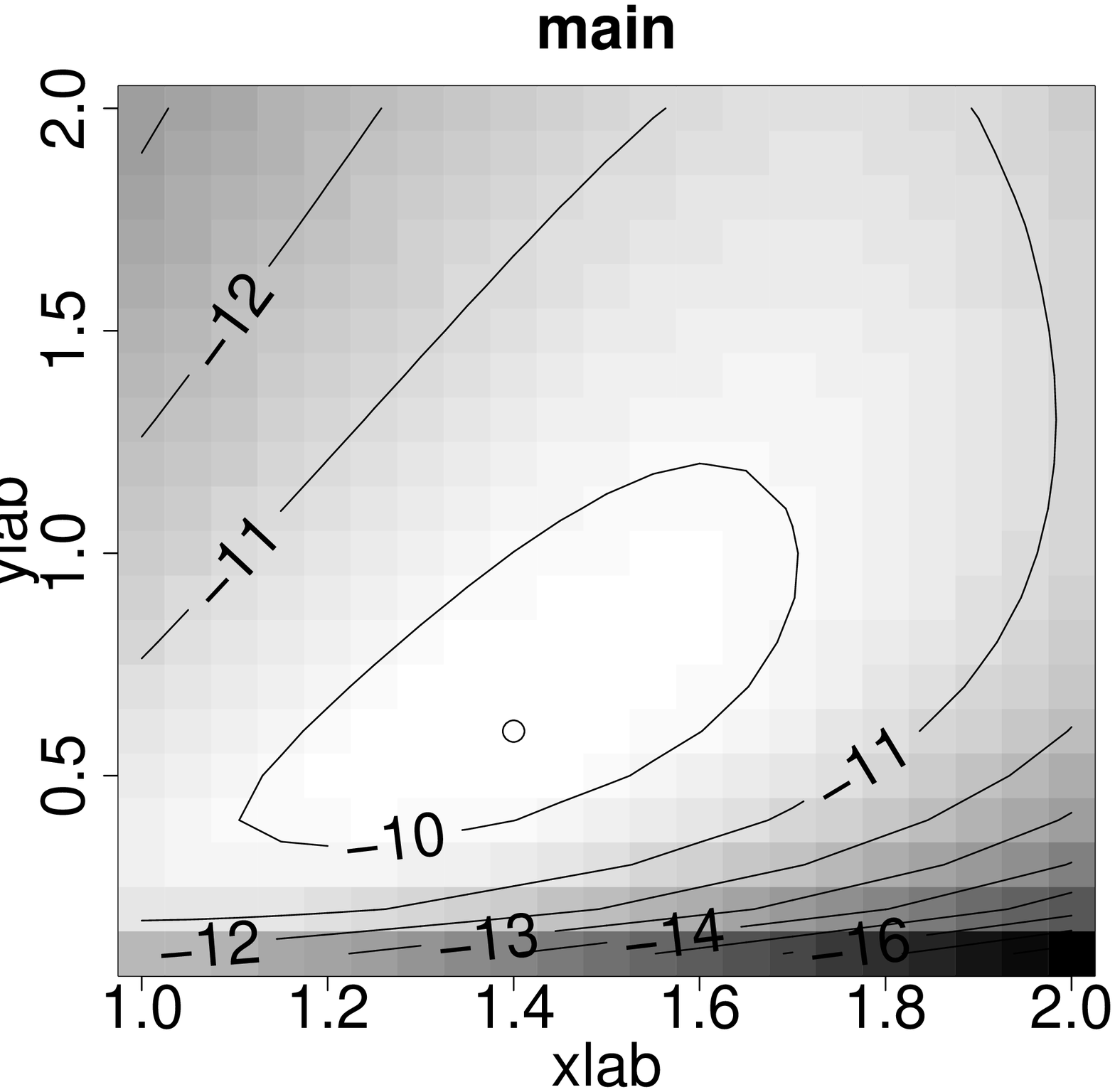}
		}\\[2ex]
		\subfigure[\cite{SA03}]{
				\label{unrooted_likelihood-surface7}
				\psfrag{main}{}
				\psfrag{xlab}{\small $\alpha$}
				\psfrag{ylab}{\small $r$}
				\includegraphics[width=40mm, height=40mm]{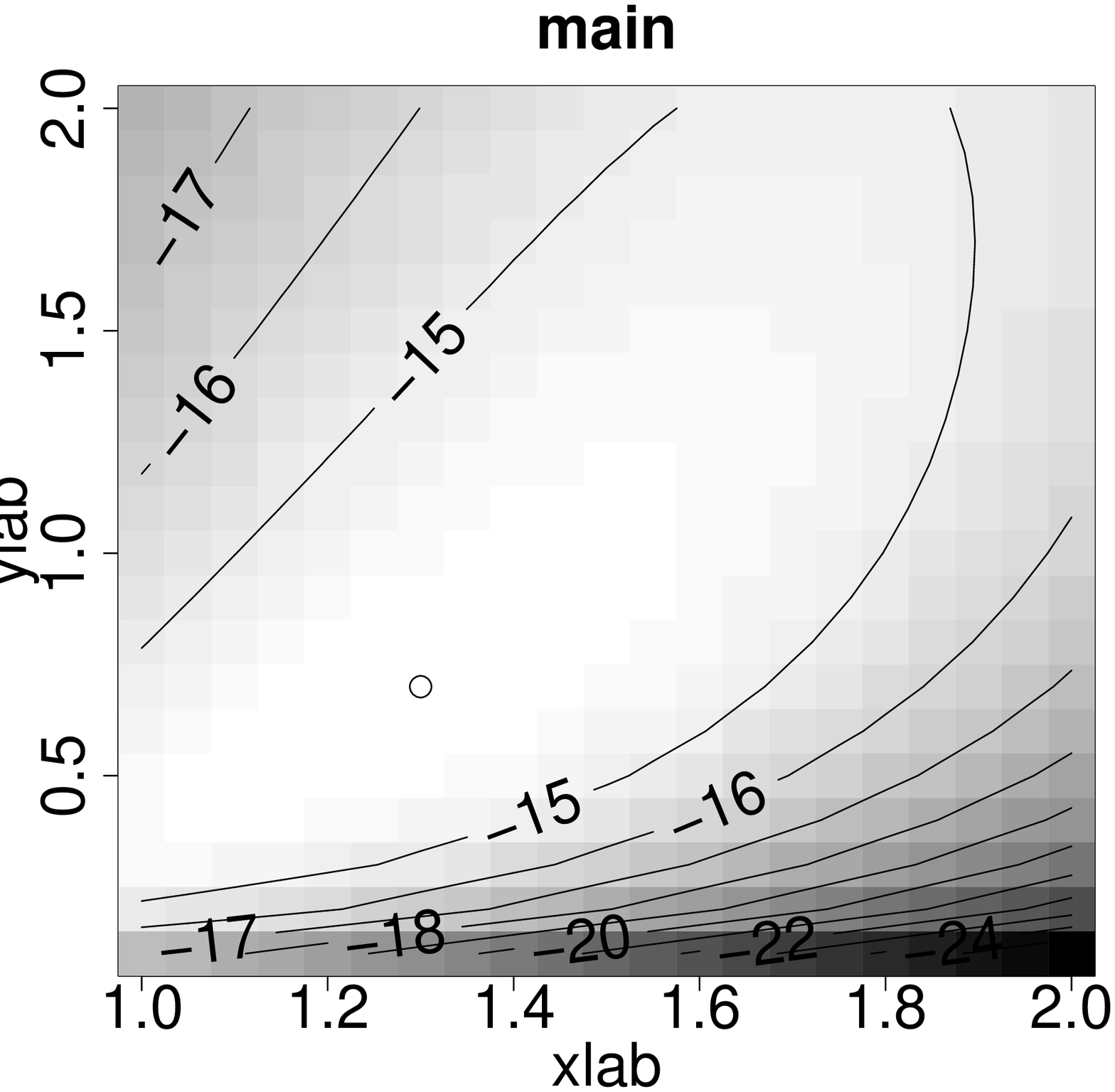}
		}
		\caption{Log${}_{10}$-Likelihood surfaces for 
unrooted cod datasets. The argmax is indicated by a dot.}\bigskip
		\label{unrooted_likelihood-surfaces}
\end{figure}

\end{document}